\documentclass[fleqn,usenatbib]{mnras}

\usepackage[T1]{fontenc}
\usepackage{ae,aecompl}

\usepackage{graphicx}	
\usepackage{amsmath}	
\usepackage{amssymb}	
\usepackage{hyperref}

\usepackage[T1]{fontenc}
\usepackage{ae,aecompl}

\usepackage{xcolor}

\title[Pulsar observation schedule and glitch measurement]{Effects of periodicity in observation scheduling on parameter estimation of pulsar glitches}

\author[L. Dunn et al.]{
    L. Dunn,$^{1,2}$\thanks{E-mail: liamd@student.unimelb.edu.au}
    M. E. Lower$^{3,4}$
    and A. Melatos$^{1,2}$
\\
$^{1}$School of Physics, University of Melbourne, Parkville, VIC 3010, Australia\\
$^{2}$Australian Research Council Centre of Excellence for Gravitational Wave Discovery (OzGrav), University of Melbourne,\\
Parkville, VIC 3010, Australia\\
$^{3}$Centre for Astrophysics and Supercomputing, Swinburne University of Technology, PO Box 218, Hawthorn VIC 3122, Australia\\
$^{4}$CSIRO Astronomy and Space Science, Australia Telescope National Facility, Epping NSW 1710, Australia
}

\date{Accepted XXX. Received YYY; in original form ZZZ}

\pubyear{2020}

\begin{document}
\label{firstpage}
\pagerange{\pageref{firstpage}--\pageref{lastpage}}
\maketitle

\begin{abstract}
In certain pulsar timing experiments, where observations are scheduled approximately periodically (e.g. daily), timing models with significantly different frequencies (including but not limited to glitch models with different frequency increments) return near-equivalent timing residuals. 
The average scheduling aperiodicity divided by the phase error due to time-of-arrival uncertainties is a useful indicator of when the degeneracy is important.
Synthetic data are used to explore the effect of this degeneracy systematically.
It is found that phase-coherent \textsc{tempo2} or \textsc{temponest}-based approaches are biased sometimes toward reporting small glitch sizes regardless of the true glitch size.
Local estimates of the spin frequency alleviate this bias.
A hidden Markov model is free from bias towards small glitches and announces explicitly the existence of multiple glitch solutions but sometimes fails to recover the correct glitch size.
Two glitches in the UTMOST public data release are re-assessed, one in PSR J1709$-$4429 at MJD 58178 and the other in PSR J1452$-$6036 at MJD 58600.
The estimated fractional frequency jump in PSR J1709$-$4429 is revised upward from $\Delta f/f = (54.6\pm 1.0) \times 10^{-9}$ to $\Delta f/f = (2432.2 \pm 0.1) \times 10^{-9}$ with the aid of additional data from the Parkes radio telescope. 
We find that the available UTMOST data for PSR J1452$-$6036 are consistent with $\Delta f/f = 270 \times 10^{-9} + N/(fT)$ with $N = 0,1,2$, where $T \approx 1\,\text{sidereal day}$ is the observation scheduling period. Data from the Parkes radio telescope can be included, and the $N = 0$ case is selected unambiguously with a combined dataset.
\end{abstract}




\section{Introduction}
The long-term spin-down of a radio pulsar may occasionally be interrupted by a glitch: an event in which the pulsar's spin frequency suddenly increases.
Glitches are typically recognised by their influence on the timing residuals \citep{espinozaStudy315Glitches2011,yuDetection107Glitches2013}, which are the deviations between the expected and measured pulse times of arrival (ToAs).
The expected arrival times are predicted by a timing model, which parametrizes both the intrinsic evolution of the rotational phase of the pulsar (frequency and frequency derivatives, glitches) as well as astrometric effects (Roemer, Shapiro, and Einstein delays, dispersion, position, proper motion, and parallax) \citep{lorimerHandbookPulsarAstronomy2004,edwardsTEMPO2NewPulsar2006}.

To estimate the parameters of a glitch, a model of the effect of the glitch on the rotational phase of the pulsar is assumed.
A typical simple form based on a Taylor expansion is \citep{lowerUTMOSTPulsarTiming2020a} \begin{equation} \Delta\phi_\mathrm{g}(t) = \Delta\phi + \Delta f(t - t_\mathrm{g}) + \frac{1}{2}\Delta\dot{f}(t-t_\mathrm{g})^2 + \ldots,\label{eqn:glitch_phase}\end{equation}
where $\Delta\phi_\mathrm{g}(t)$ denotes the extra phase accumulated in response to the glitch relative to a no-glitch phase model.
The free parameters here are $t_\mathrm{g}$, the glitch epoch; $\Delta\phi$, the permanent jump in rotational phase due to unmodelled effects or an uncertain glitch epoch; $\Delta f$, the permanent jump in spin frequency; and $\Delta\dot{f}$, the permanent jump in the first spin frequency derivative with respect to time.
These parameters may be estimated in the same way as other parameters in the pulsar timing model: through a least-squares fit which minimizes the  $\chi^2$ of the post-fit residuals \citep{hobbsTEMPO2NewPulsartiming2006}.
Alternatively they may be estimated through Bayesian inference with a software package such as \textsc{temponest} \citep{lentatiTemponestBayesianApproach2014}, which incorporates parameters describing the deterministic timing model and noise sources and calculates a posterior probability distribution for these parameters via nested sampling with \textsc{multinest} \citep{ferozMultiNestEfficientRobust2009}.
When reporting glitch parameters estimated in this way, it is tacitly assumed that the $\chi^2$ of the post-fit residuals has a unique minimum.

In this paper we explore the validity of the single-minimum assumption and the consequences for glitch parameter estimation, when the scheduling of ToA measurements is periodic.
By periodic we mean that the gap between consecutive ToAs is approximately equal to an integer multiple of some common period, e.g. if timing data are collected at the same local sidereal time for each observation.
The role of observational scheduling on pulsar glitch measurement has received some attention previously.
It is well understood that a higher density of observations is advantageous when trying to detect and characterise glitches \citep{wongObservationsSeriesSix2001,dodsonHighTimeResolution2002, janssen30GlitchesSlow2006, ashtonRotationalEvolutionVela2019, basuObservedGlitchesEight2020}, but quantitative statements along these lines are rare, due to the large number of factors which may be considered.
\citet{espinozaNeutronStarGlitches2014} pointed out that an infrequent observing cadence, combined with a jump in first frequency derivative, can mask a glitch with sufficiently small permanent frequency jumps.
They gave a quantitative lower bound on the detectable permanent frequency jump, $\Delta f_\mathrm{lim} = \Delta T\lvert\Delta\dot{f}\rvert/2$ where $\Delta T$ is the average time between observations.
Similarly, \citet{shannonCharacterizingRotationalIrregularities2016} noted in their analysis of PSR J0835$-$4510 that glitches with $\Delta f$ smaller than $10^{-7}\,\mathrm{Hz}$ are indistinguishable from timing noise, using timing data with a monthly observing cadence.
\citet{melatosPulsarGlitchDetection2020} showed that the ability of a hidden Markov model (HMM) to detect glitches with $\Delta f = 10^{-8}\,\mathrm{Hz}$ in the presence of moderate timing noise is diminished, if the time between observations exceeds $\sim10\,\mathrm{days}$.
In their timing of the frequently glitching pulsar PSR J0537$-$6910, \citet{marshallBigGlitcherRotation2004} employed a strategy in which observations are spaced logarithmically, in order to keep phase uncertainty below $0.1$ cycles without expending an inordinate amount of observing time.
Finally, the Canadian Hydrogen Intensity Mapping Experiment (CHIME) collaboration has recently noted that their determinations of the pulse frequency of newly discovered pulsars may be in error by $n/(1\,\text{sidereal day})$ where $n$ is a small integer, due to the transit nature of the instrument \citep{goodFirstDiscoveryNew2020}.
This is another manifestation of the underlying issue we explore here in detail.
To our knowledge, the specific issue of the effect of periodicity in observation scheduling on the estimation of glitch parameters has not been considered previously. 

The paper is structured as follows.
In Section~\ref{sec:conditions} we derive a practical condition for when periodic scheduling leads to errors in estimating the spin frequency of the pulsar, i.e. the time derivative of the phase to leading order in a Taylor expansion.
We demonstrate that the condition is satisfied in some existing pulsar timing datasets, leading to a near-degeneracy between timing models with significantly different spin frequencies.
In Section~\ref{sec:glitches} we specialise to the case of degeneracy between glitch models with different permanent frequency jumps.
In Section \ref{sec:glitch_param_est} we investigate how degeneracy in timing models affects specific glitch parameter estimation methods.
Section \ref{subsec:tempo2} discusses glitch parameter estimation with \textsc{tempo2} \citep{hobbsTEMPO2NewPulsartiming2006}, both in the context of phase-coherent timing (Section \ref{subsubsec:phase_coh_timing}) and local estimation of the pulsar's spin frequency (Section \ref{subsubsec:local_f0}).
Section \ref{subsec:temponest} discusses glitch parameter estimation with \textsc{temponest}, and Section \ref{subsec:hmm} discusses an HMM-based approach \citep{melatosPulsarGlitchDetection2020}.
Finally, in Section~\ref{sec:utmost} we discuss periodic scheduling in the context of the UTMOST public data release \citep{lowerUTMOSTPulsarTiming2020a}.

\section{Phase ambiguity}
\label{sec:conditions}
In this section we explore a simple, non-glitch case in which there may be ambiguity in the measurement of the evolution of the pulsar's rotational phase.
The ambiguity is ultimately due to near-periodicity in the scheduling of the ToA measurements used to determine the timing model parameters.
The case without a glitch builds intuition for the more general case with a glitch, which is discussed in Section \ref{sec:glitches}.

\subsection{Quasiperiodic scheduling}
We denote the measured ToAs by $\{t_1, t_2, \ldots, t_{N}\}$, and consider the time gap between the $i$th and $(i+1)$th ToA, $\Delta t_i = t_{i+1} - t_i$.
We call a sequence of observations ``periodic'', if we have the following approximate equality for all $i$: \begin{equation} \Delta t_i \approx n_i T, \label{eqn:obs_condition} \end{equation} where $n_i$ is an integer and $T$ is a common, fundamental period independent of $i$.

Equation (\ref{eqn:obs_condition}) expresses the periodicity condition intuitively as an approximate equality between the time gap between consecutive ToA measurements, $\Delta t_i$, and an integer multiple of some common period $T$.
However, it is more accurate to think of this condition as a restriction on the fractional part of the factor which multiplies $T$ in the exact version of equation (\ref{eqn:obs_condition}), which is not strictly an integer.
That is, we have the exact equality \begin{equation} \Delta t_i = (n_i + \epsilon_i)T,\label{eqn:obs_condition_precise}\end{equation} where $n_i$ is an integer, and the remainder $\epsilon_i$ is a real number satisfying $\lvert\epsilon_i\rvert < 0.5$ by construction.


\subsection{Phase error}
\label{subsec:phase_error}
In the limiting case of perfect periodicity, there is a choice of $T$ for which we have $\epsilon_i = 0$ for all $i$.
The degeneracy is exact: any extra term in the timing model which contributes an integer number of pulsar rotations over the time-scale $T$ predicts ToAs which coincide exactly with those measured, so timing models with and without such an extra term cannot be distinguished.
The simplest extra term is of the form \begin{equation} \Delta \phi(t) = \frac{Nt}{T}, \label{eqn:Delta_phi_simple} \end{equation} where $N$ is an integer.
Equation (\ref{eqn:Delta_phi_simple}) corresponds to a permanent change in the pulse frequency of size $\Delta f = N/T$.

The limiting case $\epsilon_i = 0$ is special.
We now allow $\epsilon_i \neq 0$, start with the true phase model $\phi(t)$, and add a spurious term of the form equation (\ref{eqn:Delta_phi_simple}).
A term of this form produces indistinguishable timing models for $\epsilon_i = 0$, so one expects the degeneracy to break slightly, when $\epsilon_i$ is small.
The degeneracy can be quantified by asking: what is the effect of the spurious phase term on the timing residuals, as a function of $\epsilon_i$?
In general the total phase residual at the $i$th ToA is given by $R_i = \phi_i - n_i$, where $\phi_i$ is the predicted pulsar rotational phase at the $i$th ToA, and $n_i$ is the closest integer to $\phi_i$ \citep{taylorPulsarTimingRelativistic1992}.
The phase residual contributed across a gap of length $\Delta t_i = (n_i + \epsilon_i) T$ by $\Delta\phi(t)$ is given by \begin{equation}\delta\phi_i = \mathrm{frac}\left[\Delta\phi(\Delta t_i)\right] = \mathrm{frac}\left[\frac{N}{T}(n_i+\epsilon_i)T\right] = \mathrm{frac}(N\epsilon_i),\label{eqn:induced_residuals}\end{equation} where $\mathrm{frac}(\cdot)$ denotes the fractional part, and the last equality follows from the fact that $N$ and $n_i$ are both integers by construction, so their product has no fractional part.
For $N=1$ we then simply have $\delta\phi_i = \epsilon_i$.

\subsection{Worked example}
\label{subsec:noglitch_worked_example}
We illustrate the arguments of this section with an example drawn from a real dataset.
The UTMOST pulsar timing programme \citep{bailesUTMOSTHybridDigital2017} uses data taken at the Molonglo Observatory Synthesis Telescope, and has released public data for a large number of pulsars \citep{lowerUTMOSTPulsarTiming2020a}.
Here we consider the dataset released for PSR J1452$-$6036, consisting of 287 ToAs measured between July 2017 and July 2019.
Our aim is to determine whether the observations of this pulsar are close enough to periodic that there is ambiguity in the phase evolution of the pulsar.
We therefore seek to determine the value of $T$ that minimises the average value of $\lvert\epsilon_i\rvert$ as defined in equation (\ref{eqn:obs_condition_precise}).
If there is a choice of $T$ which makes $\lvert\epsilon_i\rvert$ particularly small, then the phase residuals induced by a phase term of the form in equation (\ref{eqn:Delta_phi_simple}) will be correspondingly small.

\begin{figure}
    \centering
    \includegraphics[width=\columnwidth]{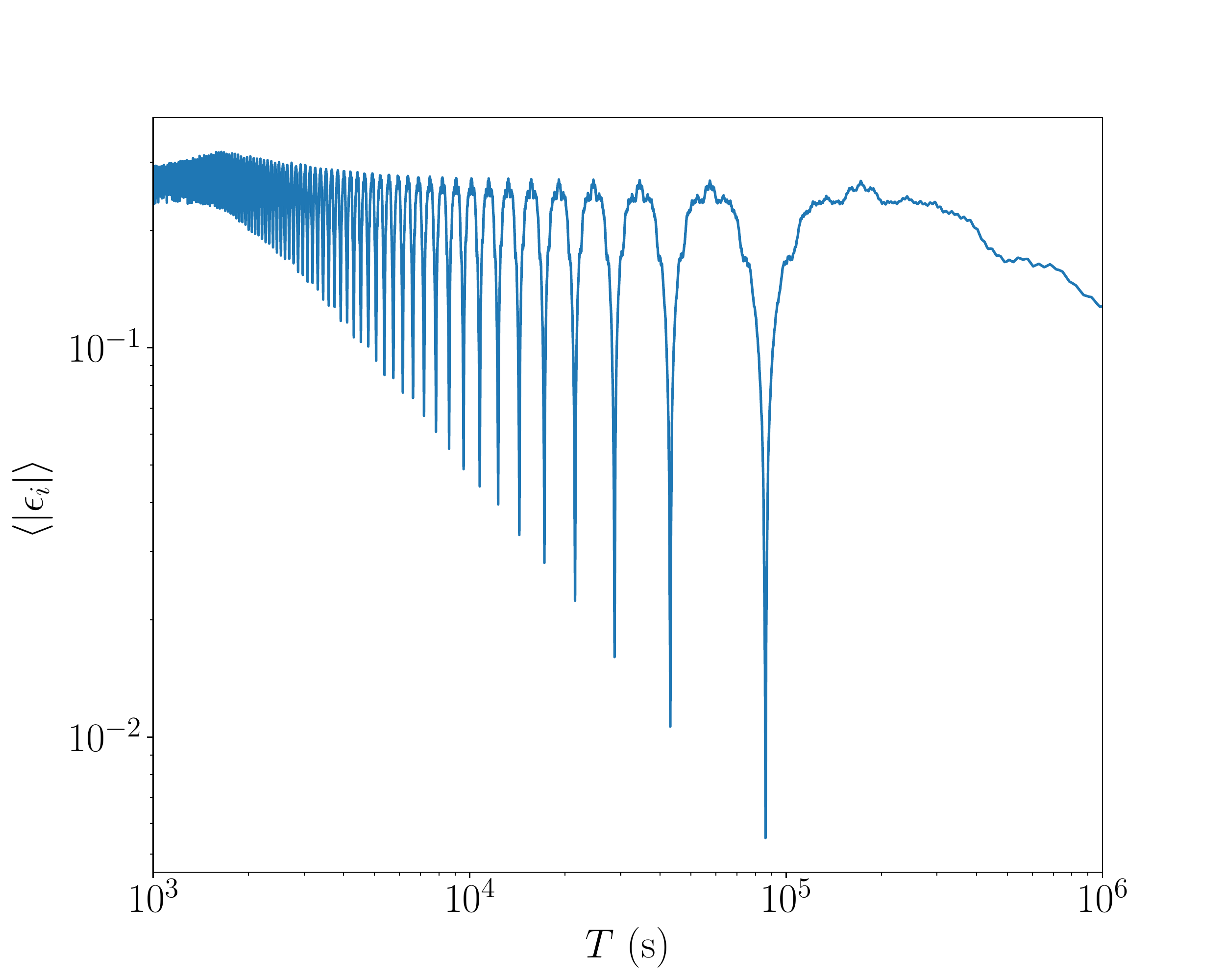}
    \caption{Mean fractional scheduling aperiodicity $\langle\lvert\epsilon_i\rvert\rangle$ as a function of scheduling quasiperiod $T$ for the UTMOST observations of PSR J1452$-$6036.}
    \label{fig:avg_epsilon}
\end{figure}

Fig. \ref{fig:avg_epsilon} graphs the average of $\lvert\epsilon_i\rvert$ as a function of $T$ for this dataset.
A sequence of local minima is clearly visible, with the lowest value of $\langle\lvert\epsilon_i\rvert\rangle$ occurring at $T = 86158\,\mathrm{s}$, and $\langle\lvert\epsilon_i\rvert\rangle = 4\times 10^{-3}$.
We therefore expect that the magnitude of the timing residuals induced by an extra phase term of the form $\Delta \phi(t) = t/(86158\,\mathrm{s})$ will be $\sim 4 \times 10^{-3}$.

\begin{figure}
    \centering
    \includegraphics[width=\columnwidth]{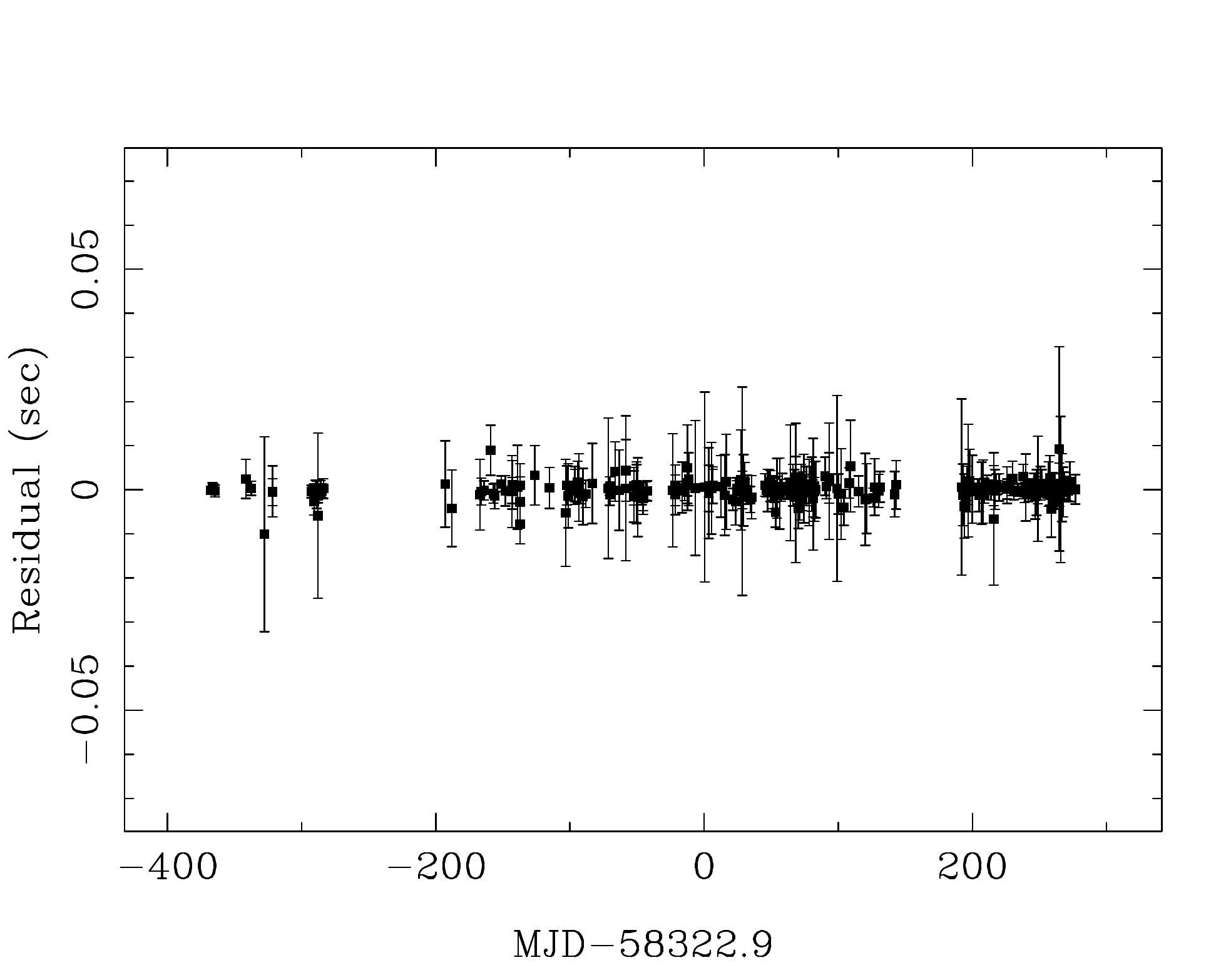}
    \includegraphics[width=\columnwidth]{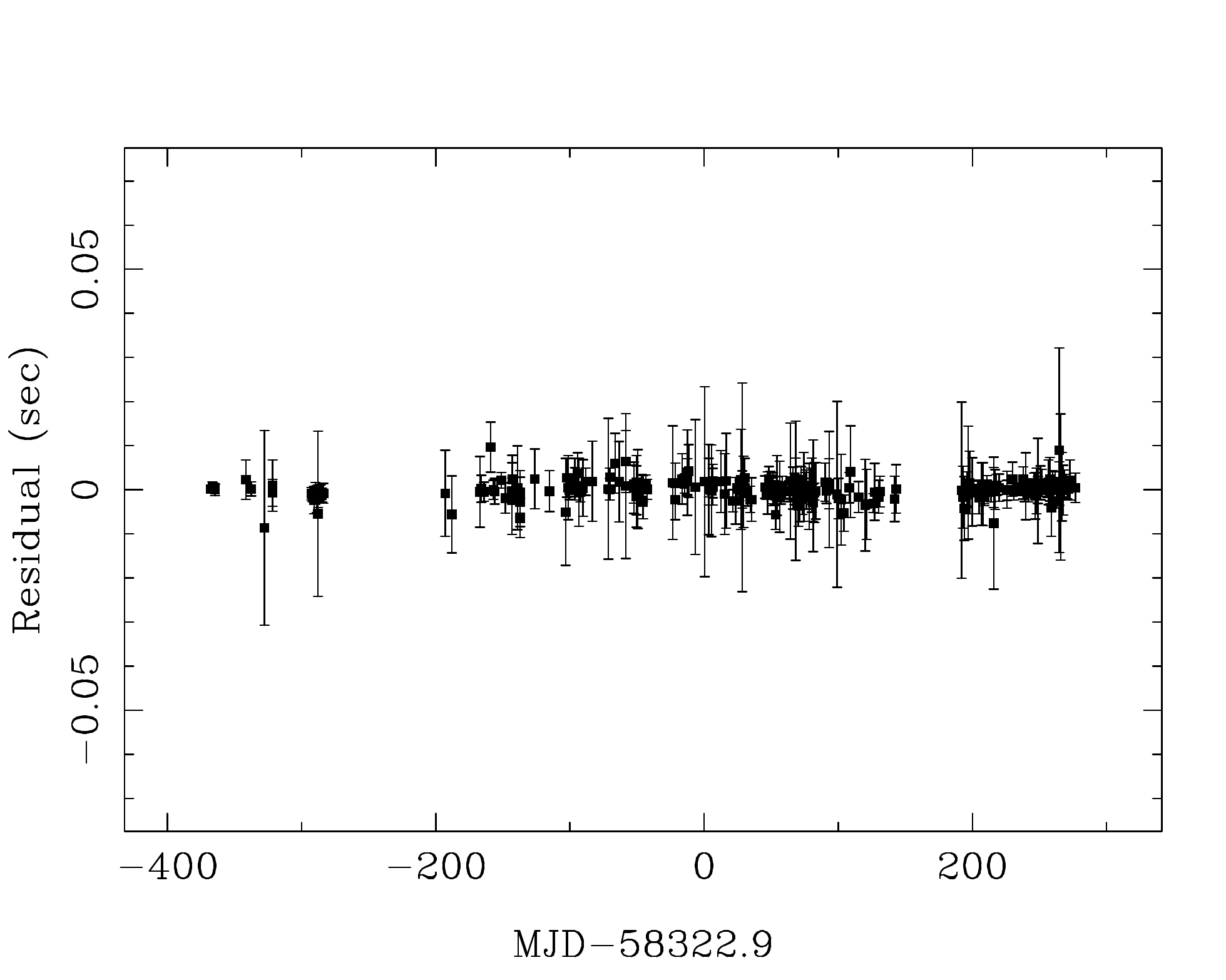}
    \includegraphics[width=\columnwidth]{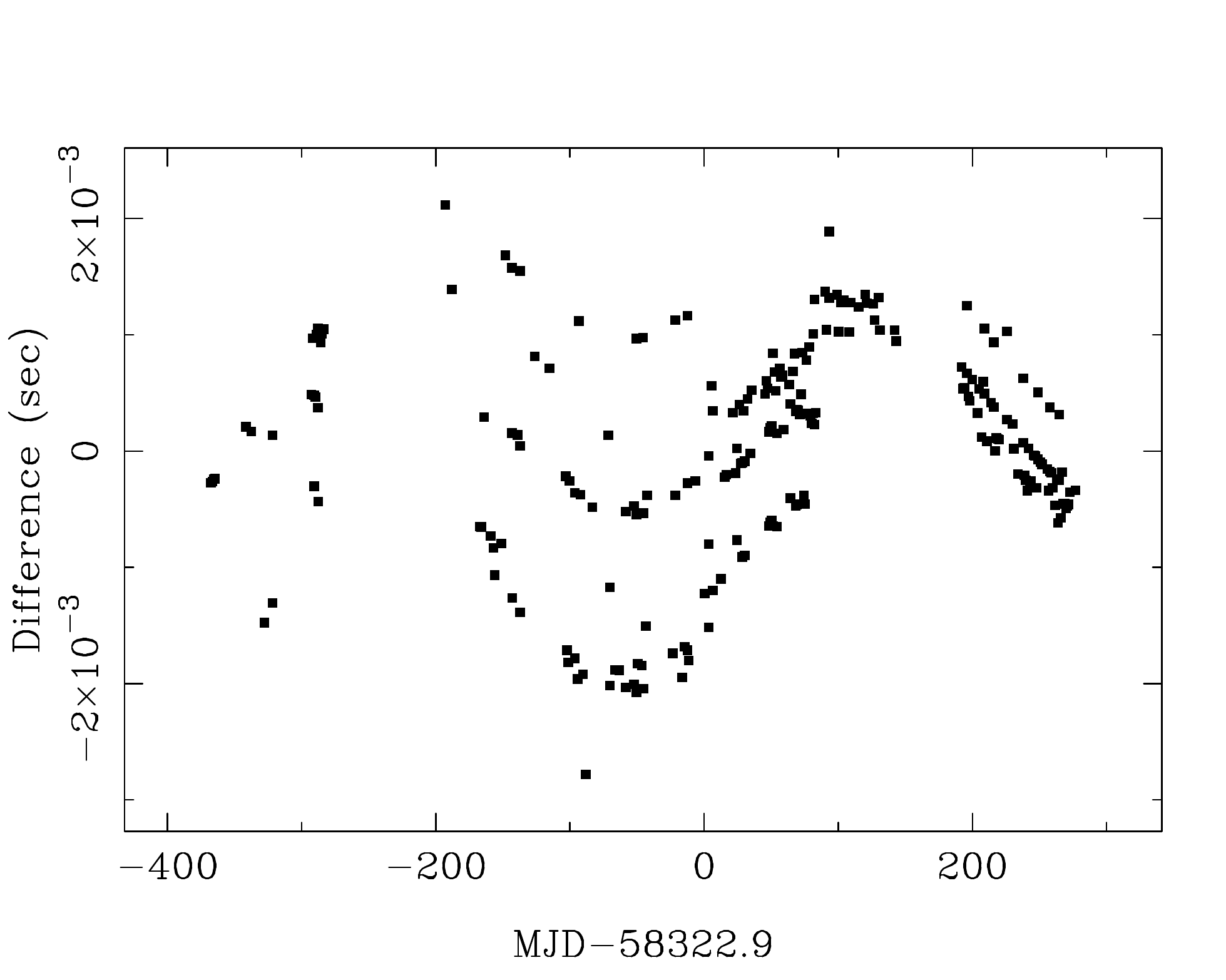}
    \caption{Timing residuals for the UTMOST observations of PSR J1452$-$6036 between MJD 57955 and MJD 58600, using the UTMOST timing model \emph{(top)}, and a model which matches the UTMOST timing model except for an overall frequency increment of $1.1605 \times 10^{-5}\,\mathrm{Hz}$ \emph{(middle)}.
    The bottom panel shows the difference between the two sets of residuals (error bars not shown).
    The uncertainty on the spin frequency reported by \textsc{tempo2} is $6 \times 10^{-10}\,\mathrm{Hz}$.}
    \label{fig:J1452-6036_UTMOST}
\end{figure}

Once we know the magnitude of the residuals induced in a dataset due to a spurious phase term of the form in equation (\ref{eqn:Delta_phi_simple}), we can compare it against the magnitude of the stochastic residuals due to ToA measurement error.
The quoted uncertainties on the ToA measurements in the PSR J1452$-$6036 data are typically $1\,\mathrm{ms}$.
The pulsar spins at roughly $6\,\mathrm{Hz}$, so ToA uncertainties $\sim1\,\mathrm{ms}$ correspond to phase residuals $\sigma_\text{ToA}\sim 6 \times 10^{-3}$.
Hence residuals induced by the extra phase term $\Delta\phi(t)$ are of the same order as the residuals due to ToA measurement error.
It is therefore reasonable to expect that the two sets of timing residuals look similar, with and without $\Delta\phi(t)$.
Equivalently, we refer to the indicative ratio \begin{equation}R = \langle\lvert\epsilon_i\rvert\rangle/\langle\sigma_{\text{ToA}}\rangle. \label{eqn:indicative_ratio}\end{equation}
If the condition $R \lesssim 1$ is satisfied, as it is in this example, we expect phase models with and without the $\Delta\phi(t)$ term to give similar residuals.
This is demonstrated in Fig. \ref{fig:J1452-6036_UTMOST}, where the top panel shows the timing residuals for the original UTMOST timing model and the middle panel shows the timing residuals for a timing model with an extra phase term $\Delta\phi(t) = t/(86158\,\mathrm{s})$ added.
The bottom panel shows the difference between the two sets of residuals.
The difference is no more than a few milliseconds, of the same order as the typical ToA error.
After re-fitting the spin frequency, the new timing model has a spin frequency $1.1605 \times 10^{-5}\,\mathrm{Hz}$ larger than the UTMOST timing model.
Presented with the top or bottom panel in Fig. \ref{fig:J1452-6036_UTMOST}, an analyst would have no reason to doubt either timing model, even though they are different.
We emphasise that in this case the degeneracy between the two timing models does not involve glitches.

\subsection{Detecting residuals from frequency misestimation}
\label{subsec:detecting}
If the frequency in the timing model is incorrect by a multiple of the inverse of the scheduling period $T$, the induced phase residual across a given ToA gap is given by equation (\ref{eqn:induced_residuals}).
We may therefore seek to detect the ``signal vector'' $\mathbfit{S}$, the $k$th element of which is the cumulative sum of the induced phase residuals up to and including the $k$th ToA gap: \begin{equation} S_k = \sum_{i=1}^k \delta\phi_i \label{eqn:error:detection_signal}.\end{equation}
In the absence of any other sources of timing residuals, $\mathbfit{S}$ is exactly what we expect to see if the frequency is misestimated due to periodic observation scheduling.
Successful detection of this signal in the residuals enables disambiguation between timing models which produce timing residuals which appear by eye to be indistinguishable.

Our treatment of this detection problem follows \citet{levyPrinciplesSignalDetection2008}.
Since we know precisely the form of the expected signal, cross-correlating the vector $\mathbfit{S}$ with the residuals $\mathbfit{R} = (R_1, R_2, \ldots, R_N)$ is an optimal means of detection if the noise in $\mathbfit{R}$ is additive, white, and gaussian.
We assume that the noise has the covariance matrix $\mathbfss{C} = \mathrm{diag}(\sigma_1^2, \sigma_2^2, \ldots, \sigma_N^2)$ where $\sigma_i$ is the reported uncertainty on the $i$th ToA, measured in cycles.
After calculating the expected signal $\mathbfit{S}$ according to equation (\ref{eqn:error:detection_signal}), we may compute the cross-correlation test statistic \begin{equation} \gamma = \mathbfit{R}^\mathrm{T} \mathbfss{C}^{-1}  \mathbfit{S} \label{eqn:error_detection_statistic}\end{equation} and compare this test statistic to a threshold $\gamma_\mathrm{th}$ chosen to give a particular false alarm rate $P_\text{fa}$, \begin{equation} \gamma_\mathrm{th} = \sqrt{\mathbfit{S}^\mathrm{T}\mathbfss{C}^{-1}\mathbfit{S}}Q^{-1}(P_\text{fa}), \label{eqn:error_detection_thresh} \end{equation} where $Q^{-1}{(x)}$ is the inverse of \begin{equation} Q(x) = \frac{1}{\sqrt{2\pi}}\int_x^\infty\mathrm{d}u\,\exp(-u^2/2). \end{equation}
The probability of detection $P_\text{d}$ for a given probability of false alarm $P_\text{fa}$ is then given by \begin{equation} P_\text{d} = 1 - Q\left[\sqrt{\mathbfit{S}^\text{T}\mathbfss{C}^{-1}\mathbfit{S}} - Q^{-1}(P_\text{fa})\right].\label{eqn:error_detection_PD}\end{equation}
This gives a quantitative estimate of the reliability of detecting the residuals caused by misestimating the frequency when observations are periodically scheduled.

For concreteness, consider again the UTMOST dataset for PSR J1452$-$6036.
We calculate $\epsilon_i$ according to (\ref{eqn:obs_condition_precise}) assuming $T = 86158\,\mathrm{s}$, and subsequently calculate $\mathbfit{S}$ assuming $N = 1$.
The latter quantity is the induced phase error signal due to the frequency in the timing model being $1/(86158\,\mathrm{s}) = 1.161 \times 10^{-5}\,\mathrm{Hz}$ too large.
For $P_\text{fa}$ fixed at $0.01$, we find $P_\text{d} = 0.98$.
Hence it is possible to reliably detect the presence of the signal $\mathbfit{S}$ due to an incorrect frequency measurement and therefore reject a timing model which contains such a signal, if one is aware of the effect and has a sufficiently long stretch of data.
We caution that this is an idealised treatment which ignores the presence of other noise sources (e.g. timing noise). If significant timing noise is present in the data, the assumption of a diagonal noise covariance matrix breaks down, and a modified approach is recommended. We note also that for all of the results presented in this section, $N$ can be either positive or negative --- it makes no difference whether the spurious phase term $\Delta\phi(t)$ correponds to an increase or decrease in frequency.

\section{Glitches}
\label{sec:glitches}

We now turn to the subject of degeneracy between glitch models.
The idea is essentially the same: under the conditions described in Section \ref{sec:conditions}, certain glitch sizes are difficult to distinguish from one another on the basis of ToA measurements, especially based on visual inspection of the timing residuals.
We begin by guiding the reader step-by-step through a worked example based on simulated data in Section \ref{subsec:glitch_worked_example} in order to illustrate the application of the key ideas.
We then show in Section \ref{subsec:freq_deriv} that a jump in the frequency derivative during a glitch does not affect the argument in Section \ref{sec:conditions} nor the conclusions in the rest of the paper.

\subsection{Worked example}
\label{subsec:glitch_worked_example}
We examine a simulated dataset that is generated as follows.
First, a timing model is chosen.
We choose to take as our starting point the timing model from the UTMOST data release for PSR J1452$-$6036, in which a glitch was detected at MJD $58600.29$ with $\Delta f = 1.745 \times 10^{-6}\,\mathrm{Hz}$, $\Delta\phi = 0$, and no reported $\Delta\dot{f}$ or exponentially recovering component \citep{lowerUTMOSTPulsarTiming2020a}.
The timing model used to generate the simulated dataset matches the timing model for PSR J1452$-$6036 from the UTMOST data release, except that the glitch at MJD $58600.29$ has size $\Delta f = 1.2106 \times 10^{-5} \,\mathrm{Hz}$ instead. All other glitch parameters are identical -- there is no phase jump at the glitch epoch, change in frequency derivative, or exponential recovery included in the simulated data.
We wish to generate a set of synthetic ToAs which are consistent with the chosen timing model.
Starting with the real PSR J1452$-$6036 dataset used in Section \ref{sec:conditions}, we use \textsc{libstempo}\footnote{\url{https://vallis.github.io/libstempo/}} to generate a new set of idealised ToAs.
The synthetic ToAs begin as an exact copy of the PSR J1452$-$6036 ToAs but are shifted slightly so that they show zero residuals when analysed with the chosen timing model.
White, gaussian noise is then added to each ToA at a level commensurate with the reported uncertainty for that ToA.
This synthetic dataset serves two purposes.
In the remainder of this section it is used to illustrate the principle that periodicity in observation scheduling leads to a degeneracy between glitch models.
In sections \ref{subsec:tempo2}, \ref{subsec:temponest}, and \ref{subsec:hmm} it is used to examine how glitch analyses with \textsc{tempo2}, \textsc{temponest}, or an HMM-based approach may be confounded by this degeneracy between glitch models.

\begin{figure}
    \centering
    \includegraphics[width=\columnwidth]{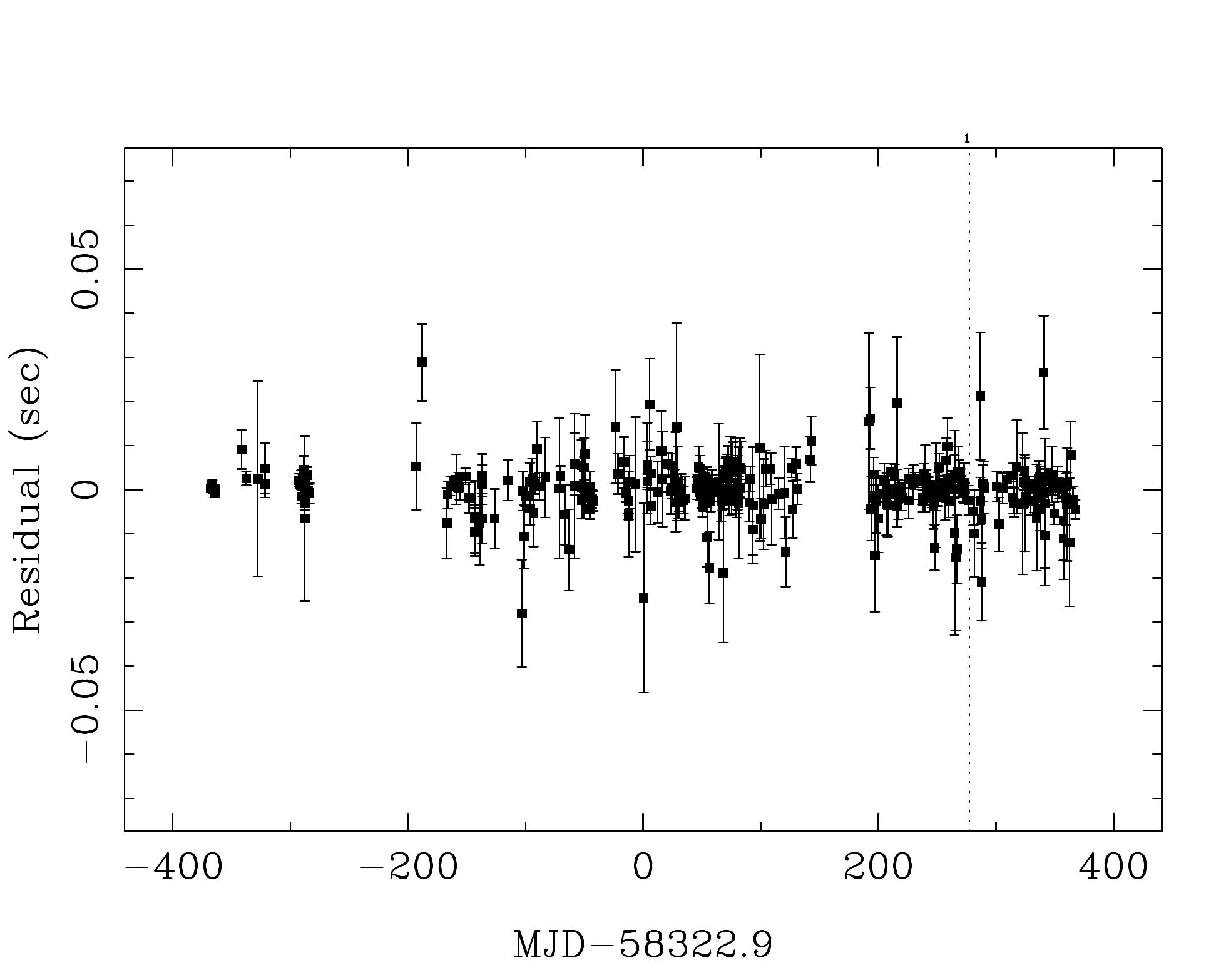}
    \includegraphics[width=\columnwidth]{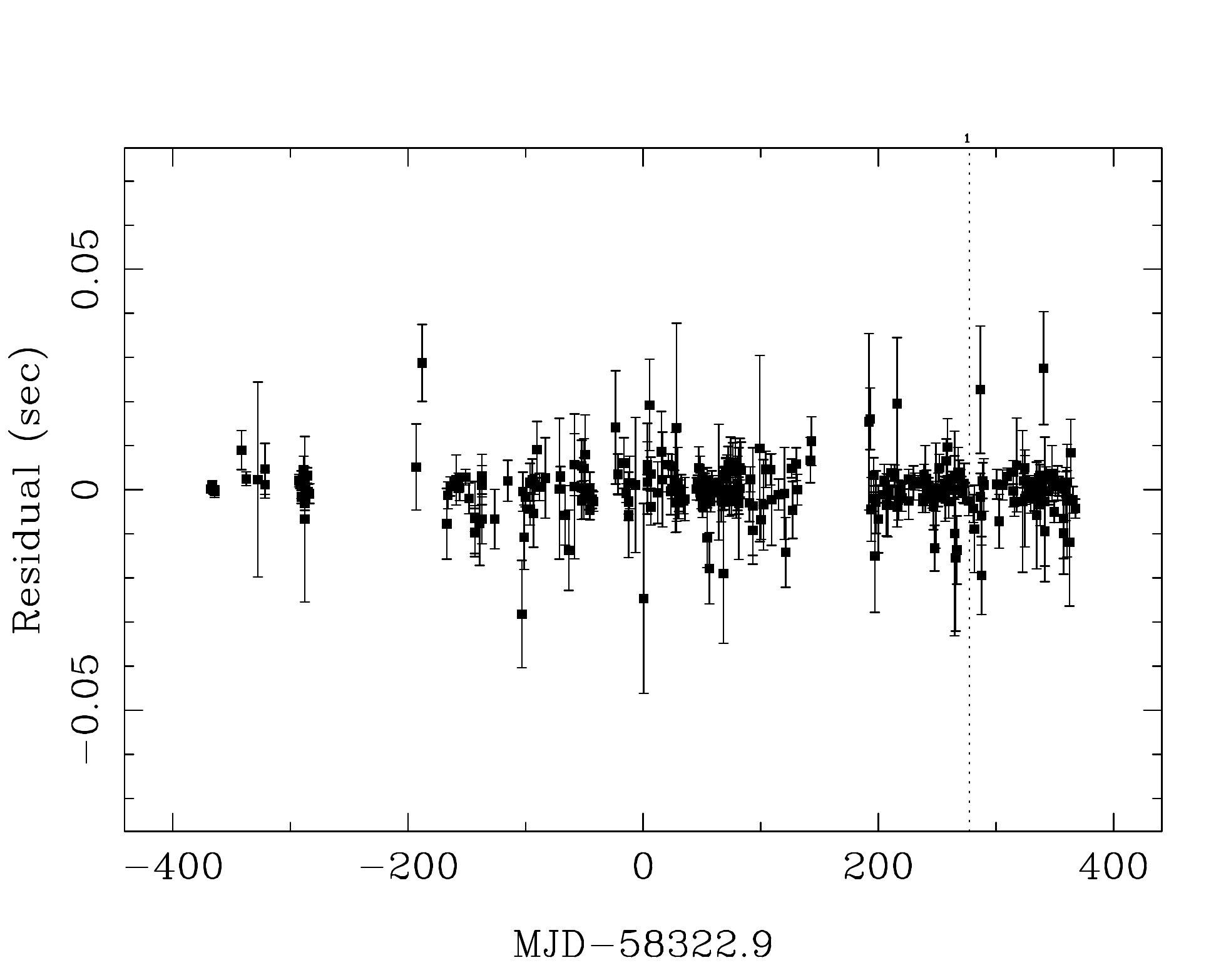}
    \caption{Timing residuals for the synthetic dataset described in Section \ref{subsec:glitch_worked_example}, using a timing model with \emph{(top)} $\Delta f = 1.2106\times 10^{-5}\,\mathrm{Hz}$  which matches the injected glitch size, and \emph{(bottom)} $\Delta f = 5 \times 10^{-7}\,\mathrm{Hz}$  which is approximately 24 times smaller than the injected glitch.}
    \label{fig:synth_J1452-6036_glitch}
\end{figure}

Fig. \ref{fig:synth_J1452-6036_glitch} compares two glitch models for the synthetic dataset described above.
The top panel of Fig. \ref{fig:synth_J1452-6036_glitch} shows the residuals for the true timing model (i.e. the chosen timing model used in generating the synthetic dataset), while the bottom panel shows the residuals for a timing model in which the glitch size is set to $5 \times 10^{-7}\,\mathrm{Hz}$.
Why $5 \times 10^{-7}\,\mathrm{Hz}$? Because the true glitch size of $1.2107\times 10^{-5}\,\mathrm{Hz}$ may be written as $1/(86158\,\mathrm{s}) + 5 \times 10^{-7}\,\mathrm{Hz}$, and hence expresses the degeneracy noted in Section \ref{sec:conditions}.
The two sets of timing residuals both appear white by eye.
The root mean square (rms) residuals for the original timing model with glitch size $\Delta f = 1.2107\times 10^{-5}\,\mathrm{Hz}$ are $2729\,\mu\mathrm{s}$, while the rms residuals for the case with a glitch size of $\Delta f = 5 \times 10^{-7}\,\mathrm{Hz}$ are $2726\,\mu\mathrm{s}$.
That is, the residuals are nearly equal, even though $\Delta f$ is approximately 24 times larger in the former model.
Note that in the glitch model with $\Delta f = 5 \times 10^{-7}\,\mathrm{Hz}$, an unphysical phase jump of $\Delta\phi = 0.326$ has also been included.

\subsection{Jumps in frequency derivative}
\label{subsec:freq_deriv}
Glitches are often accompanied by a jump in the frequency derivative, which can be a significant fraction of the pre-glitch frequency derivative.
In Section \ref{subsec:glitch_worked_example} we do not include a jump in frequency derivative, to keep things as simple as possible. 
We now show that jumps in frequency derivative do not affect the arguments made in Section \ref{sec:conditions}: the degeneracy persists even when a frequency derivative jump is present.

To be explicit, we consider two glitch models: 
\begin{align} 
    \phi_\text{g,1}(t) &= \Delta f(t - t_g) + \frac{1}{2}\Delta\dot{f}(t-t_\text{g})^2\\ 
    \phi_\text{g,2}(t) &= \left(\Delta f + \frac{1}{T}\right)(t - t_g) + \frac{1}{2}\Delta\dot{f}(t-t_\text{g})^2
\end{align}
where $\Delta f$, $\Delta\dot{f}$, and $t_\text{g}$ are defined as in equation (\ref{eqn:glitch_phase}) and $T$ is the observing period.
While these two glitch models contain a frequency derivative change, they nonetheless differ by $t/T$ (plus a constant, $t_g/T$), which has the form of equation (\ref{eqn:Delta_phi_simple}).
Therefore the arguments of Section \ref{sec:conditions} apply, and these two glitch models show comparable residuals when the observations are periodic with period $T$.

The above argument extends to glitch models with exponentially decaying terms, or any other terms which may be appropriate, as long as those other terms are common to both glitch models being compared, and the only difference between them is a frequency increment as described in Section \ref{subsec:phase_error}.

\section{Glitch parameter estimation}
In this section we discuss the effects of periodic observation scheduling on specific glitch parameter estimation techniques using \textsc{tempo2} (Section \ref{subsec:tempo2}), \textsc{temponest} (Section \ref{subsec:temponest}), and an HMM (Section \ref{subsec:hmm}).
\label{sec:glitch_param_est}
\subsection{Tempo2}
\label{subsec:tempo2}
\subsubsection{Phase-coherent timing}
\label{subsubsec:phase_coh_timing}
In Section \ref{sec:glitches} and Figure \ref{fig:synth_J1452-6036_glitch} we demonstrate that when a glitch occurs in a pulsar, there may be multiple timing models which give similarly small timing residuals.
Of course, the illustrative worked example in Section \ref{subsec:glitch_worked_example} is fashioned deliberately by exploiting our foreknowledge of the glitch parameters injected into a synthetic data set.
Does the same outcome arise ``naturally'', when the true glitch parameters are unknown, e.g. in real astronomical data or blind injections in synthetic data?
In this section we perform a phase-coherent timing analysis using \textsc{tempo2} of the same synthetic dataset.
This is not a truly blind analysis, but the steps taken in the course of this analysis approximate one typical course of action for the parameter estimation of a previously unknown glitch using astronomical data.

\begin{figure}
    \centering
    \includegraphics[width=\columnwidth]{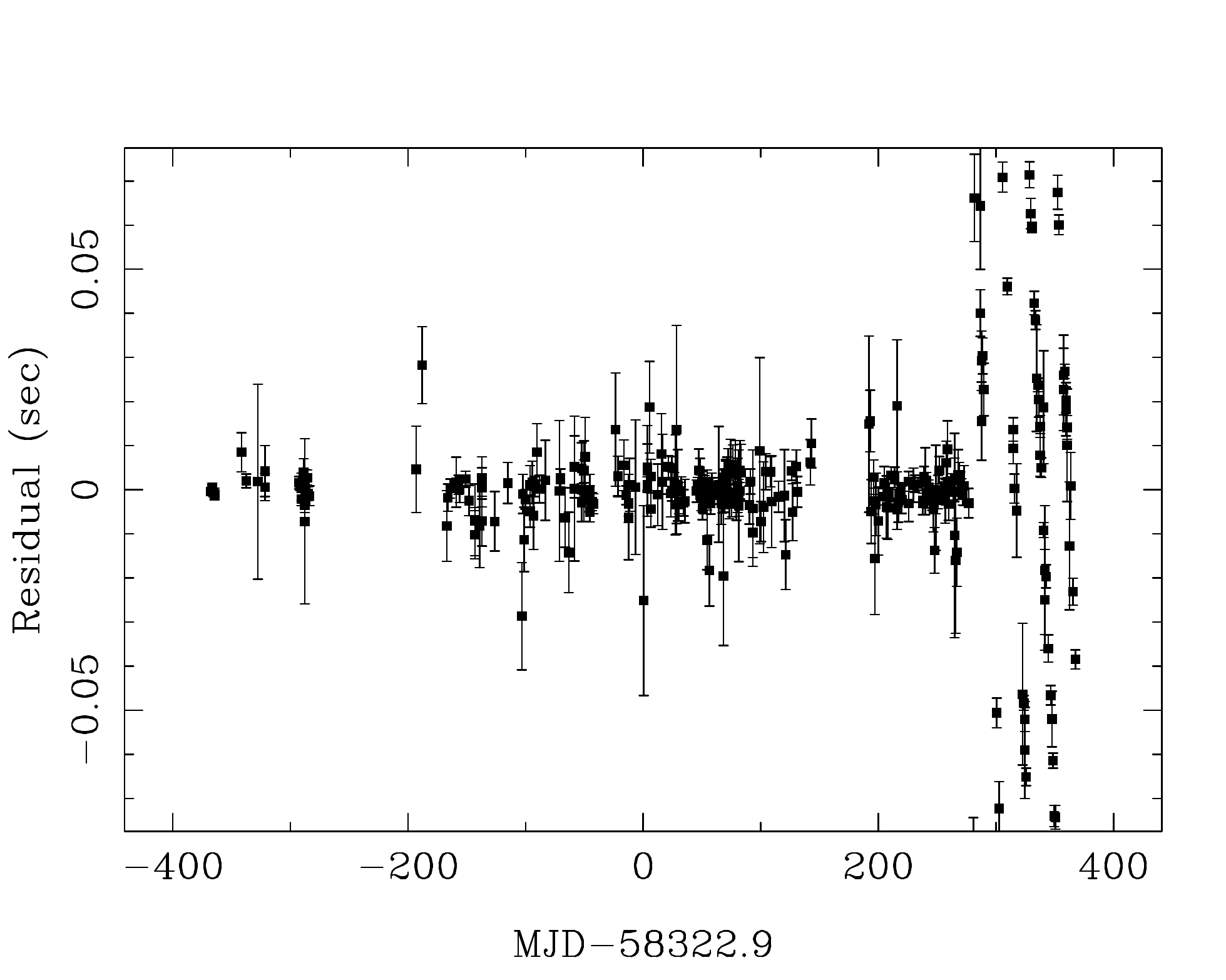}\\
    \includegraphics[width=\columnwidth]{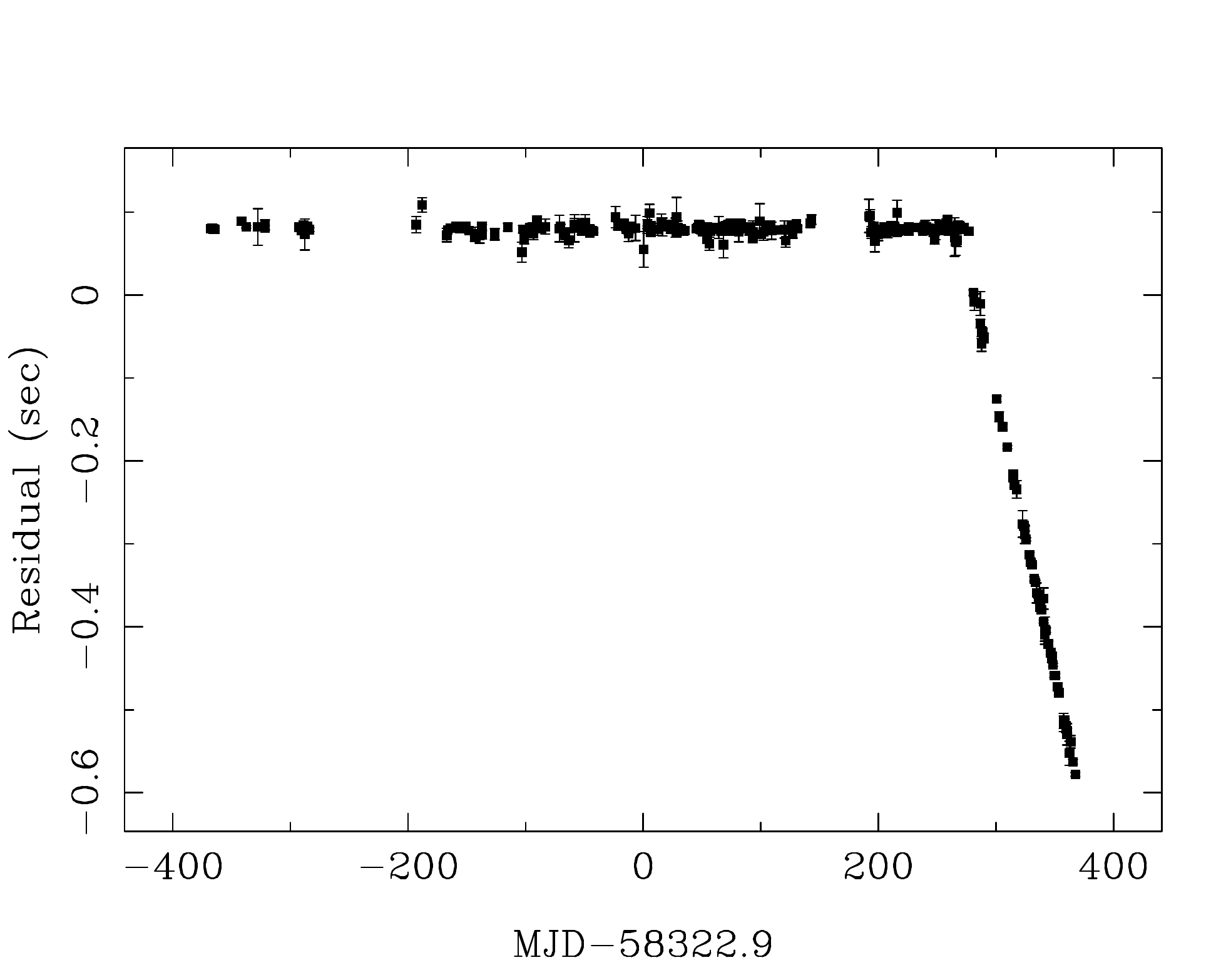}\\
    \includegraphics[width=\columnwidth]{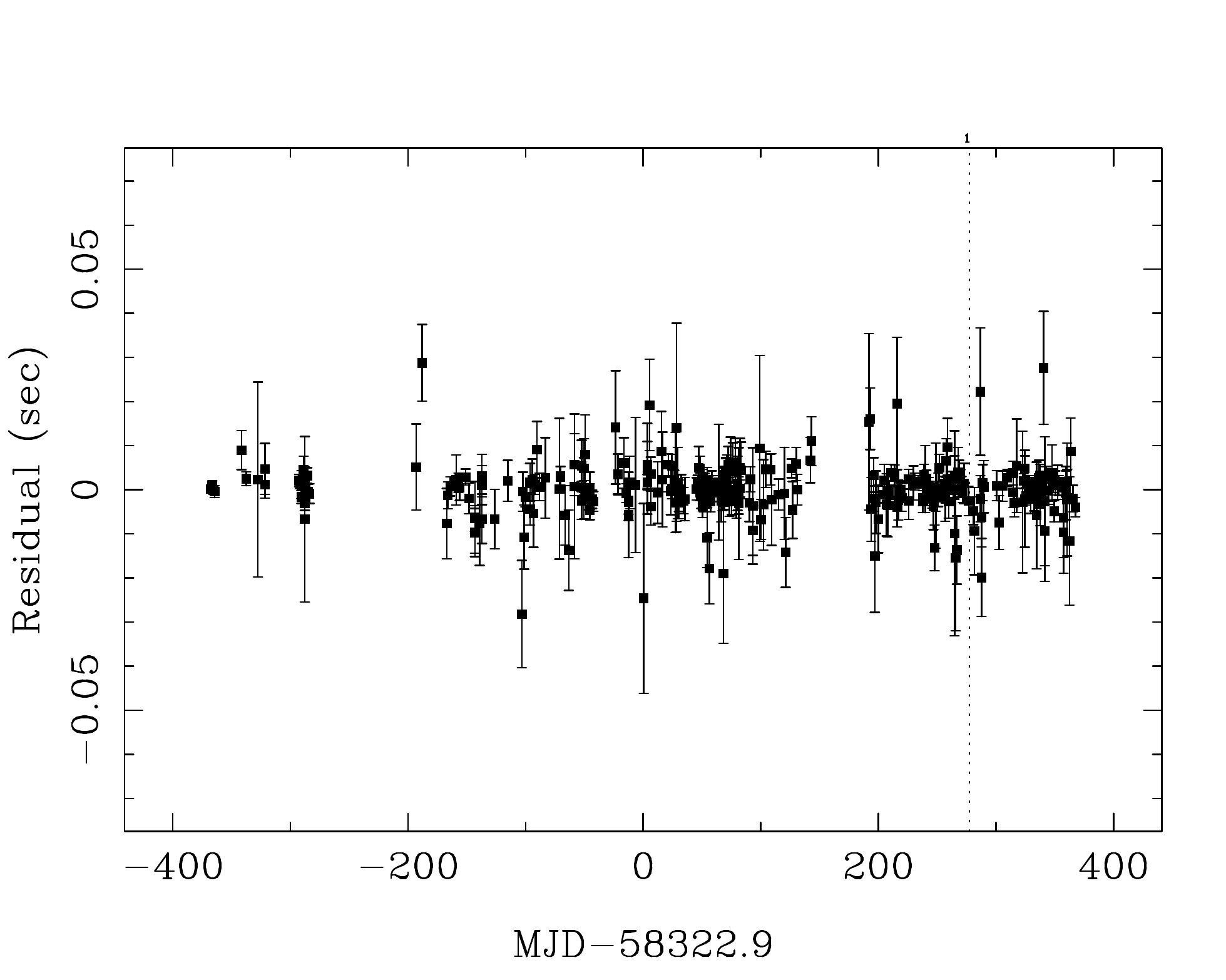}
    \caption{Timing residuals for the synthetic dataset described in Section \ref{subsec:glitch_worked_example} before (\emph{top panel}) and after (\emph{middle panel}) adding phase jumps to account for wraps in pulse phase.
    The bottom panel shows the residuals obtained after accounting for a glitch with $\Delta f = (5.006 \pm 0.015) \times 10^{-7}\,\mathrm{Hz}$ and $\Delta\phi = 0.323 \pm 0.008$.}
    \label{fig:synth_J1452-6036_tempo2}
\end{figure}

We begin by inspecting the timing residuals for the synthetic dataset used in Section \ref{subsec:glitch_worked_example} with a timing model that matches the injected parameters but without a glitch.
The residuals are shown in the top panel of Fig. \ref{fig:synth_J1452-6036_tempo2}.
The pre-glitch residuals are close to white, but there is a clear point at MJD 58600 where the timing model ``fails'', and the residuals diverge and begin to wrap.
At epochs where the phase residuals wrap around, we add phase jumps by hand to restore phase connection.
At this stage we use \textsc{tempo2} to tag each ToA with the number of pulses which have elapsed since the first ToA.
We emphasise that this pulse-numbering is model-dependent.
With the phase jumps added, we see a clear transition from flat phase residuals to a linear ramp in the residuals in the middle panel of Fig. \ref{fig:synth_J1452-6036_tempo2} -- a clean glitch signature.
Using the pulse-numbered ToAs it is straightforward to fit the glitch parameters with \textsc{tempo2}.
For simplicity we fit only the glitch frequency increment $\Delta f$ and an unphysical glitch phase jump $\Delta \phi$ to account for uncertainty in the glitch epoch.
We set the glitch epoch to be the same as the injected glitch epoch, MJD $58600.29$, for ease of comparison with the injected glitch parameters.
The \textsc{tempo2} fit returns $\Delta f = (5.008 \pm 0.006)\times 10^{-7} \,\mathrm{Hz}$, and $\Delta\phi = 0.320 \pm 0.003$.
The associated residuals are plotted in the bottom panel of Fig. \ref{fig:synth_J1452-6036_tempo2}.
By eye, they appear close to white, with variance consistent with the ToA error bars, which are of order $1\,\mathrm{ms}$.
This timing model is a good fit: the reduced $\chi^2$ returned by \textsc{tempo2} is $0.98$.
It is not, however, an accurate recovery of the injected glitch parameters, which are $\Delta f = 1.2106 \times 10^{-5}\,\mathrm{Hz}$ and $\Delta\phi = 0$.

The above analysis exemplifies a general principle: given a set of ToAs containing a glitch, if the ToAs are consistent with multiple glitch models due to periodicity in the observation schedule, then the glitch model with the smallest $\Delta f$ is more likely to be recovered by a phase-coherent timing analysis.
This occurs because ToAs which are consistent with a small glitch display a relatively gentle linear ramp in the post-glitch residuals.
This linear ramp will be picked out by eye, and subsequently used to number the post-glitch pulses.
A larger glitch (e.g. $1.2106\times 10^{-5}\,\mathrm{Hz}$ compared to $5 \times 10^{-7}\,\mathrm{Hz}$) has a different phase model, which assigns a different numbering to the post-glitch pulses. 
Once the analyst restores phase connection (whether correctly or incorrectly) and numbers the pulses according to this phase connected solution, the range of possible glitch models is restricted.

\subsubsection{Local frequency estimation}
\label{subsubsec:local_f0}
Rather than constructing a phase-connected timing solution and subsequently estimating glitch parameters, it is also possible to estimate the frequency evolution \emph{locally} by fitting for the frequency using small sets of ToAs closely spaced in time which are derived from sub-integrations of a longer observation.
In this case, the inter-ToA spacing is much smaller than $T$, the fundamental observing period.
We stipulate that the ToAs are closely spaced in time to sidestep the issue of distinguishing between phase models with frequencies differing by $1/T$, which arises once ToAs separated by more than $T$ are included in the fit.
In order to distinguish between timing models which differ in spin frequency by $1/T$, the time $\tau$ between the first and last ToA in each local fit must be long enough that the accumulated phase error exceeds the phase error $\sigma_\text{ToA}$ due to the ToA uncertainty: \begin{equation} \tau/T \gg \sigma_\text{ToA}.\end{equation}

\begin{figure}
    \centering
    \includegraphics[width=0.9\columnwidth]{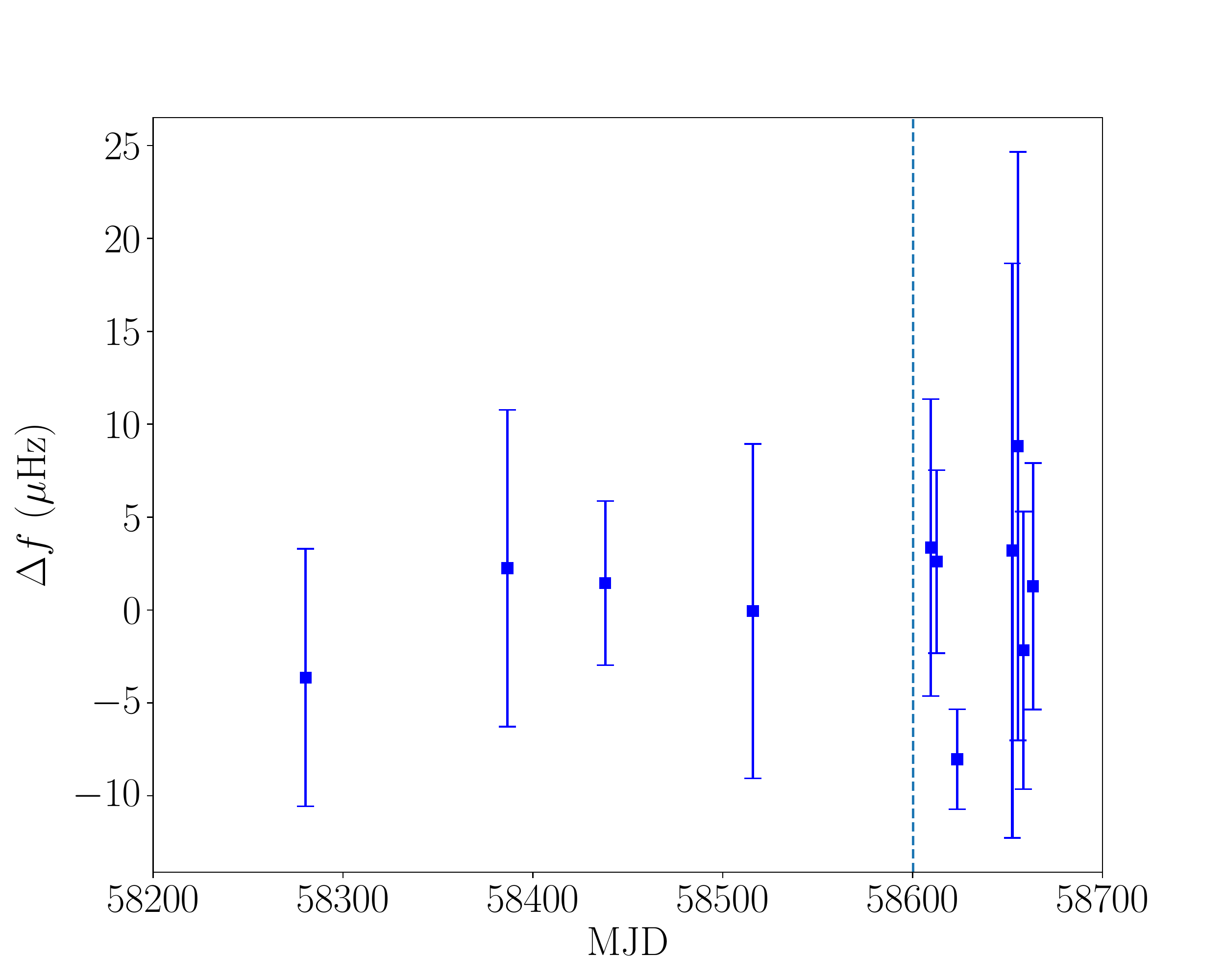}\\
    \includegraphics[width=0.9\columnwidth]{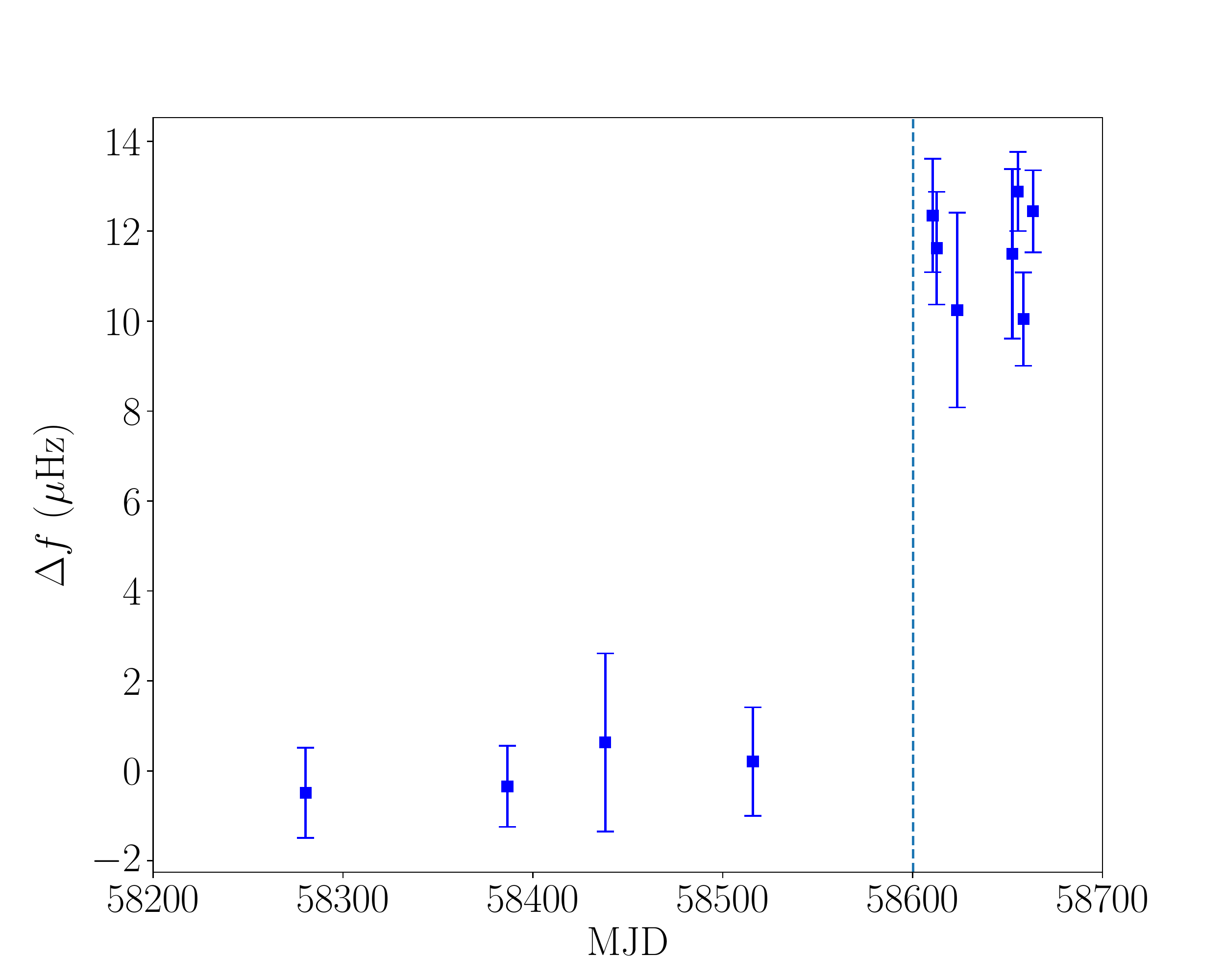}\\
    \includegraphics[width=0.9\columnwidth]{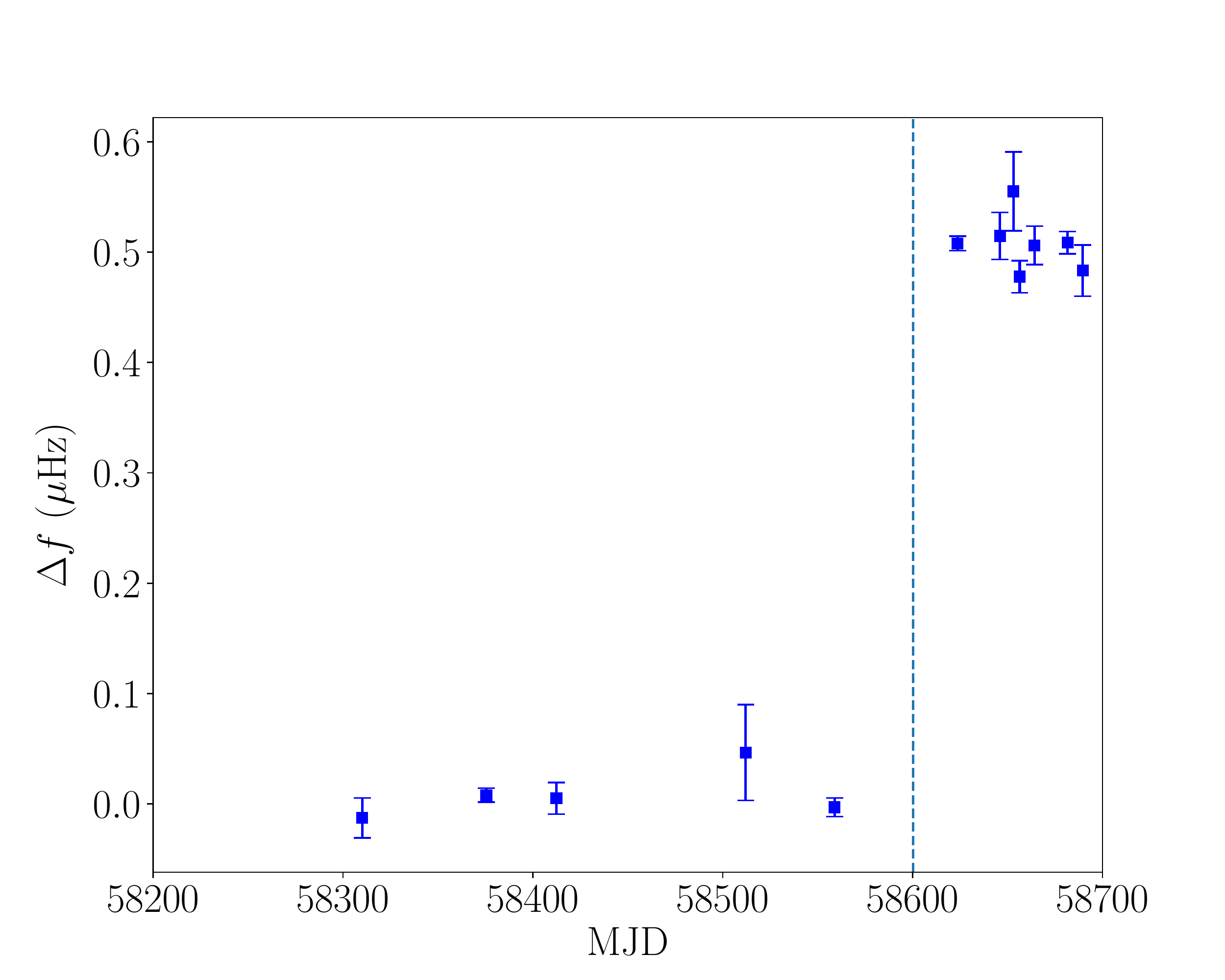}
    \caption{Results of local spin frequency estimation described in Section \ref{subsubsec:local_f0}.
    The top panel shows the case $\tau/T \approx 2\sigma_\text{ToA}$, the middle panel shows the case $\tau/T \approx 10\sigma_\text{ToA}$, and the bottom panel shows the case where multiple observing sessions are used in each spin frequency fit.
    The location of the glitch is indicated by the vertical dotted line.
    In all cases the true size of the glitch is $\Delta f = 1.21 \times 10^{-5}\,\mathrm{Hz}$.
    Note the $10^2$-fold decrease in the vertical scale from the top to the bottom panels.}
    \label{fig:local_f0}
\end{figure}

We consider three applications of local frequency estimation to synthetic data.
For each application we generate a synthetic dataset with the same timing model parameters and observation schedule as described in Section \ref{subsec:glitch_worked_example} (so the observation period $T$ is still roughly $86158\,\mathrm{s}$).
At each session when a synthetic observation is made, we generate four ToAs spaced closely, with $\tau \ll T$.
In the first of the three applications, we take $\tau = 1100\,\mathrm{s}$,\footnote{This corresponds to one ToA every $4.5$ minutes, which is quite typical for pulsars observed by UTMOST.} giving $\tau/T \approx 2\sigma_\text{ToA}$.
The results of the local spin frequency estimation for this dataset are shown in the top panel of Fig. \ref{fig:local_f0}.
The scatter in the post-glitch frequency estimates is on the order of $10\,\mu\mathrm{Hz}$, prohibiting a determination of the post-glitch spin frequency which distinguishes between glitch models with sizes separated by $1/T \approx 1.2 \times 10^{-5}\,\mathrm{Hz}$.
However, if $\tau$ is increased by a factor of $5$, we instead have the result shown in the middle panel of Fig. \ref{fig:local_f0}.
The scatter in the frequency estimation is much reduced to $\sim 1\,\mu\mathrm{Hz}$, and it is clear that a glitch with $\Delta f \approx 1.2 \times 10^{-5}\,\mathrm{Hz}$ is preferred over the value recovered by the phase-coherent analysis in Section \ref{subsubsec:phase_coh_timing} ($\Delta f \approx 5 \times 10^{-7}\,\mathrm{Hz}$).
Finally, we can analyse the same synthetic dataset but use eight ToAs spanning two consecutive observing sessions in each local frequency estimate, so that $\tau \approx T$.
The results for this case are shown in the bottom panel of Fig. \ref{fig:local_f0}.
The scatter in post-glitch frequency estimations is again reduced, down to roughly $0.1\,\mu\mathrm{Hz}$.
However, the local estimates are centred around $5 \times 10^{-7}\,\mathrm{Hz}$ --- the same incorrect estimate recovered by the phase-coherent analysis in Section \ref{subsubsec:phase_coh_timing}.
In both the phase-coherent case and the $\tau \approx T$ case, the assumption that the increase in frequency due to the glitch contributes less than one full rotation between two observation sessions leads to the incorrect estimate.

\subsection{TempoNest}
\label{subsec:temponest}
The Bayesian pulsar timing package \textsc{temponest} is sometimes used to estimate glitch parameters \citep{shannonCharacterizingRotationalIrregularities2016,yuDetectionProbabilityNeutron2017, parthasarathyTimingYoungRadio2020, lowerUTMOSTPulsarTiming2020a}.
If the glitch parameters are not already well-constrained, it is common to first use \textsc{tempo2} to obtain a set of pulse-numbered ToAs, following the procedure described in Section \ref{subsubsec:phase_coh_timing}.
Note that it is impossible to run \textsc{temponest} without pulse-numbered ToAs, unless the timing model parameters are already well-constrained.
As noted in Section \ref{subsubsec:phase_coh_timing}, pulse-numbering restricts the range of viable glitch models.
Even if one chooses an uninformative prior for $\Delta f$, \textsc{temponest} only assigns high posterior probability to glitch models which are consistent with the pulse-numbering used.
As such, glitch sizes estimated  with \textsc{temponest} exhibit the same bias towards small glitch sizes as those estimated with a phase-coherent \textsc{tempo2} analysis.


\subsection{Hidden Markov model}
\label{subsec:hmm}
Recently \citet{melatosPulsarGlitchDetection2020} presented a new approach to pulsar glitch detection that models the rotational evolution of the pulsar with an HMM and selects between models with and without glitches present. 
Like \textsc{tempo2}-based approaches, the HMM operates on ToAs.
Therefore the arguments in sections \ref{sec:conditions} and \ref{sec:glitches} suggest that one ought to be careful when estimating glitch parameters with an HMM, if the observations are scheduled periodically.
It turns out that the HMM estimates $\Delta f$ more accurately than \textsc{tempo2} and \textsc{temponest} for the synthetic dataset from Section \ref{subsec:glitch_worked_example}, as reported in Section \ref{subsubsec:hmm_example}.
However it is still somewhat prone to the same ambiguities arising from periodic scheduling, as demonstrated in Section \ref{subsubsec:hmm_ambiguity} with the aid of specific examples.

The HMM consists of three essential components, which we briefly describe here.
We refer the reader to Appendix \ref{apdx:hmm_params} and \citet{melatosPulsarGlitchDetection2020} for further details.
In the HMM, the state of the pulsar is described by its pulse frequency $f$, and the time derivative of the pulse frequency $\dot{f}$.
The values of $f$ and $\dot{f}$ are measured relative to a fiducial frequency evolution which is taken from a Taylor expansion of the phase model calculated by \textsc{tempo2}.
For example, the state $(f, \dot{f}) = (0,0)$ indicates that the state of the pulsar is exactly as predicted by the Taylor expansion at that timestep.
The state $(f, \dot{f})$ is ``hidden'': it is not observed directly, since the HMM operates only on ToAs.
Instead, we define an ``emission probability'' $L(\Delta t_i; f, \dot{f})$ which gives the probability of observing a given ToA gap $\Delta t_i$ if the pulsar's state is $(f, \dot{f})$ during this gap.
The expression for $L(\Delta t_i; f, \dot{f})$ used in this paper is given by equation (\ref{eqn:von_mises}).
Finally, it is necessary to specify the ``transition probabilities'' which determine the probability that one state transitions to another state after each ToA gap.
The HMM tracks the spin wandering directly, as a realization of a Markov chain.
This is in contrast to \textsc{temponest}, which estimates the ensemble characteristics (power spectral density) of the residuals due to timing noise \citep{lentatiTemponestBayesianApproach2014}.
In this work we use transition probabilities which assume a random walk in the second frequency derivative of the pulsar.
This matches the prescription adopted by \citet{melatosPulsarGlitchDetection2020}.
Other, qualitatively similar forms of the transition probabilities are also possible and produce qualitatively similar results.
Once detected, the parameters of the glitch may be estimated by constructing the sequence of \emph{a posteriori} most likely $(f, \dot{f})$ states of the pulsar using the forward-backward HMM algorithm, and subsequently reading off $\Delta f$ at the most probable glitch epoch.
With the parameter choices adopted in this work, the HMM is computationally cheap -- the search for glitches described in Section \ref{subsubsec:hmm_example} takes roughly $10\,\mathrm{min}$ to run on a modern desktop CPU (the quoted time is measured on an Intel Core i5-9300H CPU running at $2.40\,\mathrm{GHz}$.)

\subsubsection{Worked example}
\label{subsubsec:hmm_example}
To illustrate the effect of periodic scheduling on glitch measurement with an HMM, we re-analyse the same synthetic dataset presented in Section \ref{subsec:glitch_worked_example} and analysed with \textsc{tempo2} in Section \ref{subsec:tempo2}.
The details of the parameter choices in the HMM analysis can be found in Appendix \ref{apdx:hmm_params}.
Most relevant to the issue at hand is the range of $f$, which is taken to be $[-3 \times 10^{-7}, 2.5 \times 10^{-5}]\,\mathrm{Hz}$, bracketing the \textsc{tempo2} timing solution.
This range of frequencies is chosen to encompass a large enough range of glitch sizes to demonstrate the essential point. In particular it covers $\Delta f \pm 1/(1\,\text{sidereal day})$, where $\Delta f = 1.21 \times 10^{-5}\,\mathrm{Hz}$ is the injected glitch size. A wider frequency range is possible, but would significantly increase the computation time required without adding anything new [see Section 4.4 of \citet{melatosPulsarGlitchDetection2020}].
A glitch is detected at the 231st ToA gap, between MJDs 58599 and 58603, with log Bayes factor $250$.
We denote this model by $M_1(231)$.
While the epoch matches the injected glitch, which occurs at MJD 58600.29, we mimic a realistic analysis by persevering and searching ``blindly'' for a second glitch by comparing $M_1(231)$ to a set of two-glitch models $M_2(231, k)$ which contain glitches during the 231st gap and the $k$th gap ($k \neq 231$).
This procedure is the greedy hierarchical algorithm described in Section 4.2 in \citet{melatosPulsarGlitchDetection2020}.
We find that the two-glitch model $M_2(231, 232)$ is favored over $M_1(231)$ with a log Bayes factor of $53$.
We then compare the two-glitch model $M_2(231, 232)$ to a set of three-glitch models $M_3(231, 232, l)$.
None of the three-glitch models are favoured over $M_2(231, 232)$ with log Bayes factor less than zero for all $l \neq 231, 232$, so we terminate the search.
Note that we do not interpret $M_2(231, 232)$ as a model with two truly distinct spin-up events, because the ToA gaps are adjacent.
$M_2(231,232)$ is favoured over $M_1(231)$ because the one-glitch models are constructed such that the glitch occurs at the beginning of the ToA gap,
whereas in these data the glitch occurs nearly one day into a four-day long gap.
The two-glitch model contains enough freedom to mitigate the phase error caused by the simple form of the glitch model assumed by the HMM, which only allows for a glitch to occur at the beginning of a ToA gap, and includes no phase jumps.

\begin{figure}
    \centering
    \includegraphics[width=\columnwidth]{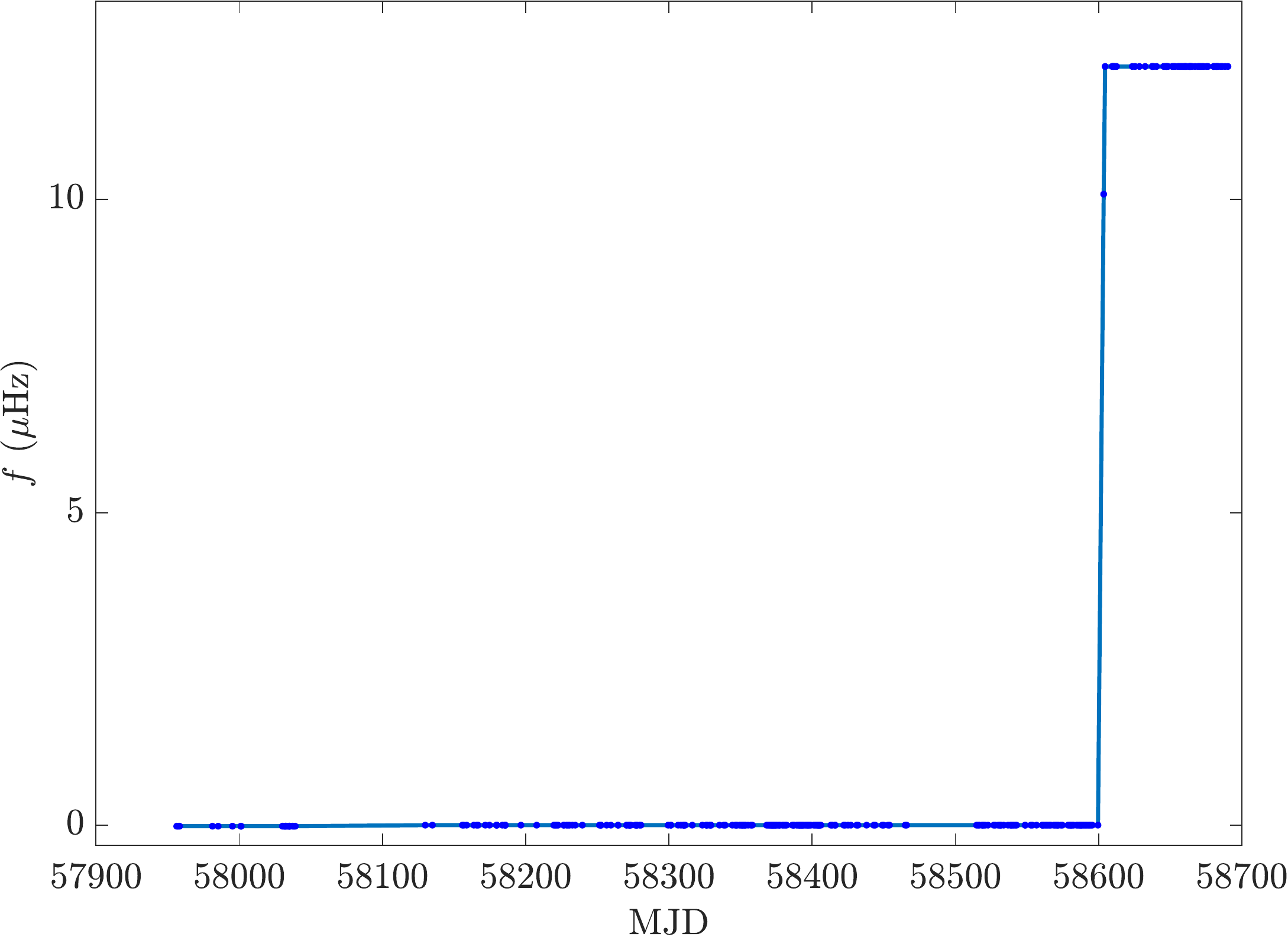}\\\vspace{0.4cm}
    \includegraphics[width=\columnwidth]{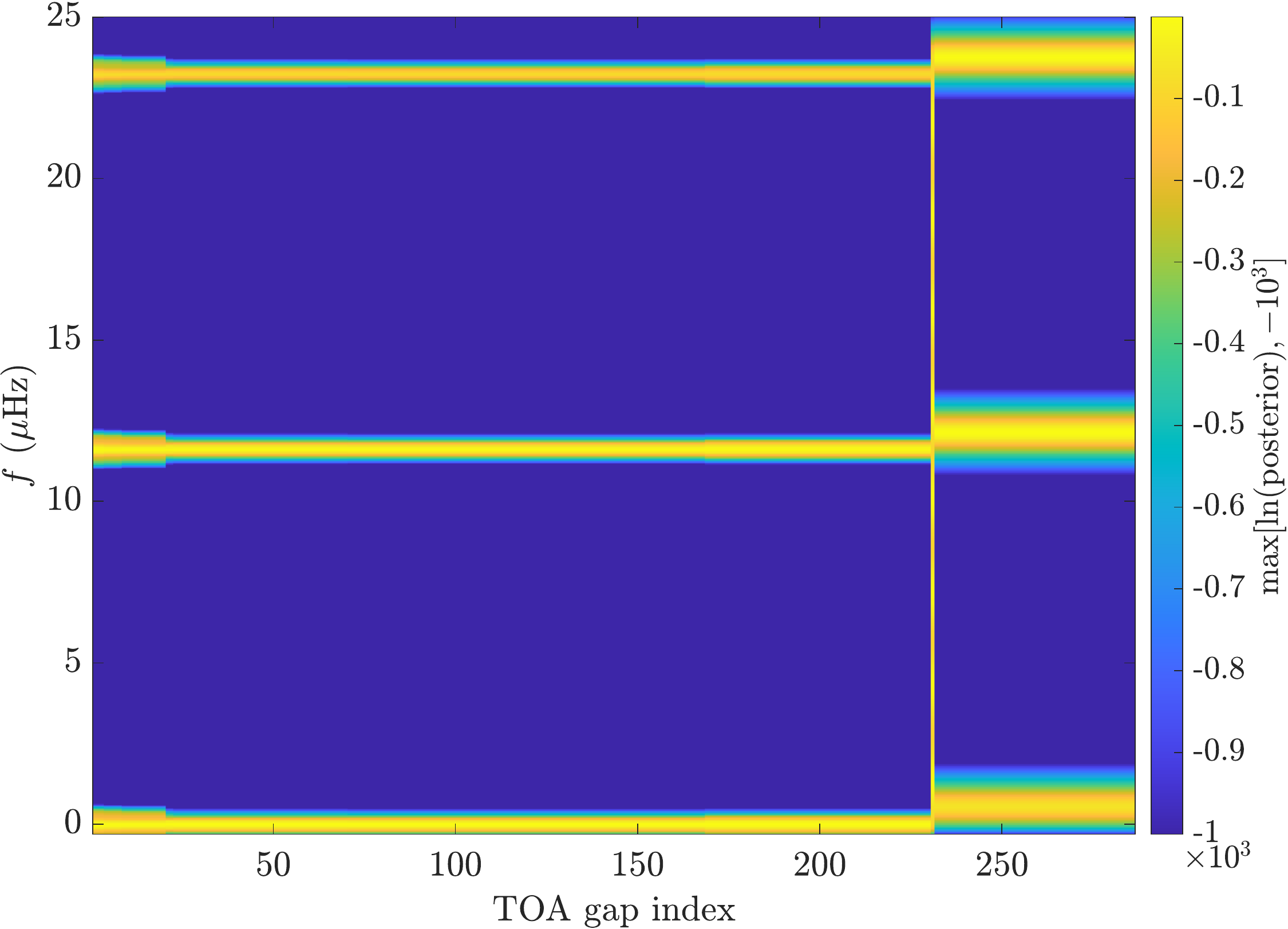}
    \caption{\emph{(Top)} Frequency path recovered by the HMM forward-backward algorithm for the analysis of the synthetic dataset described in Section \ref{subsec:glitch_worked_example}. The recovered glitch size is $\Delta f = 1.2078 \times 10^{-5}\,\mathrm{Hz}$.
    \emph{(Bottom)} Heatmap showing the evolution of the logarithm of the posterior probability for $f$ from the HMM analysis of the same synthetic dataset.
    Time increases from left to right, and at each timestep the posterior has been marginalised over $\dot{f}$.
    Values in the heatmap are clipped below $-10^3$ to aid readability.
    The favoured model, $M_2(231, 232)$, includes glitches during ToA gaps 231 and 232.}
    \label{fig:synth_J1452-6036_hmm}
\end{figure}

Given $M_2(231,232)$, the HMM forward-backward algorithm computes the posterior distribution of $f$ and $\dot{f}$ during each ToA gap. From this posterior distribution we construct a sequence of frequency and frequency derivative states by choosing the \emph{a posteriori} most likely states during each ToA gap \citep{rabinerTutorialHiddenMarkov1989,melatosPulsarGlitchDetection2020}.
The frequency sequence $f(t_i)$ constructed in this way is shown in the top panel of Fig. \ref{fig:synth_J1452-6036_hmm}.
The location of the glitch is clear.
Reading off the size of the glitch gives $\Delta f = 1.2078 \times 10^{-5}\,\mathrm{Hz}$, compared to the true glitch size of $1.2107 \times 10^{-5}\,\mathrm{Hz}$.
While the glitch model allows for a change in $\dot{f}$ (which can be positive or negative), we do not observe such a change in the recovered $\dot{f}(t_i)$ sequence.
This is in agreement with the injected glitch parameters, which include $\Delta\dot{f} = 0$.

While this appears to be a relatively successful recovery of the glitch parameters -- more so than the phase-coherent analysis of Section \ref{subsubsec:phase_coh_timing} -- we can further inspect the posterior distribution of spin states to determine what effect the observation schedule has on this mode of analysis.
A heatmap of the logarithm of the posterior distribution of $f$ (marginalised over $\dot{f}$) as a function of ToA gap index is shown in the bottom panel of Fig. \ref{fig:synth_J1452-6036_hmm}.
It exhibits a multiply peaked structure both before and after the glitch, with peaks separated by $1.1618 \times 10^{-5}\,\mathrm{Hz}$.
This spacing is significant: it is close to $1/(86158\,\mathrm{s}) = 1.1607 \times 10^{-5}\,\mathrm{Hz}$, where $86158\,\mathrm{s}$ is the observation scheduling period for this dataset.
These multiple peaks in the posterior $f$ distribution are indicative of the degeneracy caused by periodic observation scheduling.
While the peaks appear to be equal in height based on the logarithmic heatmap, after the glitch the peak at $\Delta f = 1.2078 \times 10^{-5}\,\mathrm{Hz}$ is systematically higher (by a factor of at least 1.5) than the other two, and therefore features in the recovered sequence.

\subsubsection{Robustness of parameter estimation against uncertainty in glitch epoch}
\label{subsubsec:hmm_ambiguity}
The HMM succeeds in recovering the correct glitch size in Section \ref{subsubsec:hmm_example} despite the periodic scheduling.
However, small variations in the epoch of the glitch can significantly perturb the recovered frequency path when the scheduling is periodic.

\begin{figure}
    \centering
    \includegraphics[width=\columnwidth]{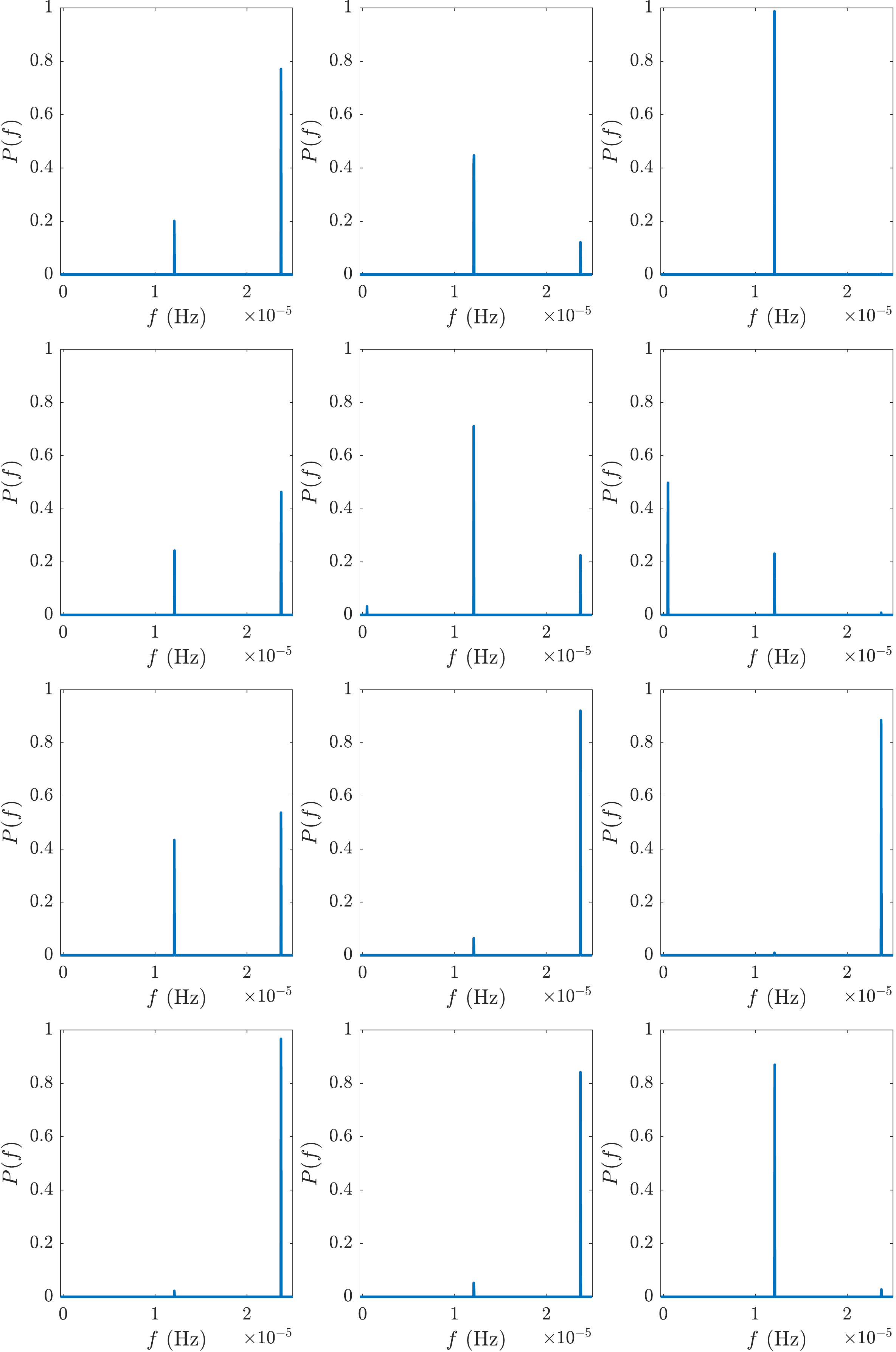}
    \caption{Posterior probability distributions $P(f)$ of $f$ during the 250th ToA gap (well after the glitch at $t_{231}$) for twelve synthetic datasets with randomised glitch epochs in the interval MJD $58599 < t_\text{g} < $ MJD $58603$ as described in Section \ref{subsec:hmm}. In all cases the true frequency deviation is $1.2107 \times 10^{-5}\,\mathrm{Hz}$ and the glitch occurs during the 231st ToA gap.}
    \label{fig:synth_J1452-6036_hmm_freq_posteriors}
\end{figure}

Fig. \ref{fig:synth_J1452-6036_hmm_freq_posteriors} shows the marginalised post-glitch posterior distributions of $f(t_{250})$ ($t_{250}$ is chosen to show the posterior distribution of $f$ well after the glitch, which occurs at $t_{231}$) for $12$ injected glitches which have the same size as the previous example, $\Delta f = 1.2107 \times 10^{-5}\,\mathrm{Hz}$, but epochs distributed uniformly between MJD $58599$ and MJD $58603$, i.e. anywhere within the 231st ToA gap.
The posterior in each case has the same multiply peaked structure as in the bottom panel of Fig. \ref{fig:synth_J1452-6036_hmm}, with the largest peak located randomly near one of three values: $5 \times 10^{-7}\,\mathrm{Hz}$, $1.2 \times 10^{-5}\,\mathrm{Hz}$, and $2.4 \times 10^{-5}\,\mathrm{Hz}$.
In all but one of the panels of Fig. \ref{fig:synth_J1452-6036_hmm_freq_posteriors}, only one or two peaks of the three-peak structure seen in the bottom panel of Fig. \ref{fig:synth_J1452-6036_hmm} are high enough to be seen by eye.

Thus HMM-based analyses of datasets with periodic scheduling do not necessarily recover the correct glitch size: of the twelve posterior $f$ distributions shown in Fig. \ref{fig:synth_J1452-6036_hmm_freq_posteriors}, only four peak at the correct location of $1.2107 \times 10^{-5}\,\mathrm{Hz}$.
The rest have peaks displaced from the true glitch size by approximately $1/(86158\,\mathrm{s})$, indicating that the ambiguity due to periodic scheduling is responsible for the failures to recover the correct glitch size in the other realisations.

Phase-coherent timing analyses of similar datasets are biased towards recovering the smallest plausible glitch size.
The HMM shows no systematic bias.
Nonetheless it is not guaranteed to return the true glitch size either.
Therefore the methods complement one another and are safest to use in tandem.
If periodic scheduling is unavoidable for some reason, a chance discrepancy between the methods is one way to catch errors in the estimate of $\Delta f$, as the example in Sections \ref{subsubsec:phase_coh_timing}, \ref{subsubsec:local_f0}, and \ref{subsubsec:hmm_example} demonstrates in practice.
Interestingly the HMM posterior distribution readily reveals the existence of multiple high-likelihood frequency tracks, which are not seen so easily in a timing analysis.

\section{Re-assessing UTMOST glitches}
\label{sec:utmost}

From mid-2017 onwards it was decided to use the Molonglo Observatory Synthesis Telescope in transit-only mode \citep{venkatramankrishnanUTMOSTSurveyMagnetars2020}.
In this mode, the transit of astronomical objects through the primary beam occurs at the same time each sidereal day, with a typical dwell time of 5--20 minutes \citep{lowerUTMOSTPulsarTiming2020a}.
Thus the ToAs used in the UTMOST timing programme are collected in clusters separated by integer multiples of $1$ sidereal day ($86164\,\mathrm{s}$), closely matching the observing period of $86158\,\mathrm{s}$ seen in the UTMOST dataset for PSR J1452$-$6036.
In most cases the cadence of observations is 3--7 days.

\begin{figure}
    \includegraphics[width=\columnwidth]{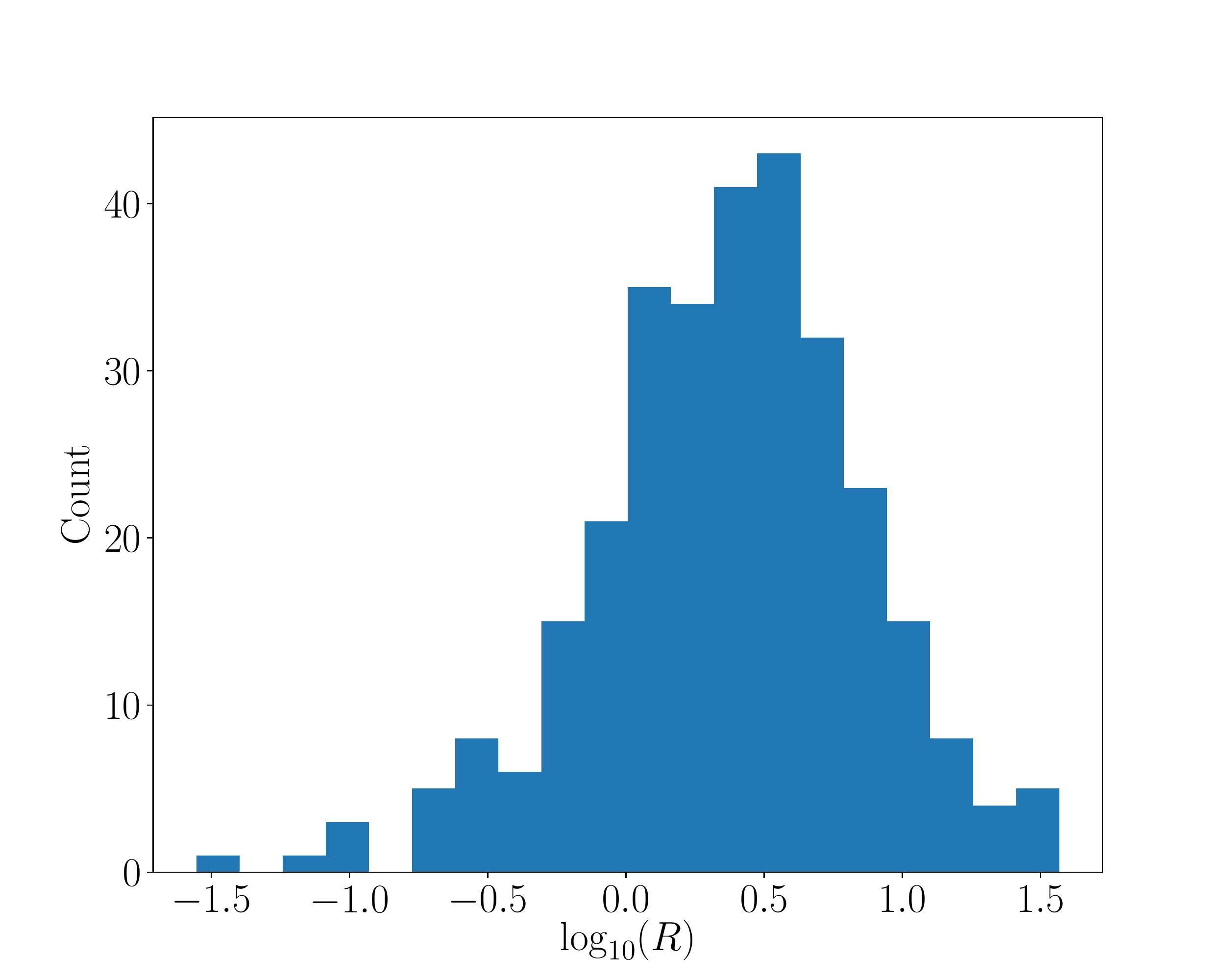}
    \caption{Histogram of $R$ for all 300 pulsars in the UTMOST data release.}
    \label{fig:UTMOST_epsilon_rms_ratios}
\end{figure}

With this in mind, we check to see if periodic scheduling compromises the phase reconstruction of UTMOST pulsars.
We calculate $\langle\lvert\epsilon_i\rvert\rangle$ [where $\epsilon_i$ is defined in equation (\ref{eqn:obs_condition_precise})] for each of the 300 pulsars in the UTMOST data release, using only ToAs measured after September 2017 in transit mode, and assuming $T = 86164\,\mathrm{s}$.
Recall that $\langle\lvert\epsilon_i\rvert\rangle$ measures the magnitude of timing residuals (measured in terms of cycles) which can be expected if one chooses a timing model which displaces $f$ by $1/T$ from its true value, where $T$ is the scheduling period.
Fig. \ref{fig:UTMOST_epsilon_rms_ratios} shows a histogram of $R$ [defined in equation (\ref{eqn:indicative_ratio})], for the 300 targets.
Datasets with $R \gtrsim 1$ are relatively unlikely to support multiple glitch models with comparable residuals.
The residuals induced by the wrong choice of $\Delta f$ are so large, that the error is obvious.
However, there is a sizeable population of pulsars with $\langle\lvert\epsilon_i\rvert\rangle/\langle\sigma_\text{ToA}\rangle < 1$, including two pulsars which are known to have glitched since September 2017: PSR J1709$-$4429 ($\langle\lvert\epsilon_i\rvert\rangle/\langle\sigma_\text{ToA}\rangle = 0.81$) and PSR J1452$-$6036 ($\langle\lvert\epsilon_i\rvert\rangle/\langle\sigma_\text{ToA}\rangle = 0.12$).
The glitch in PSR J1709$-$4429 has not been reported by any timing programmes other than UTMOST.
Given the periodic observation schedule, the possibility that the reported glitch sizes are in error by $1.16 \times 10^{-5}\,\mathrm{Hz}$ or more must be taken seriously.
We now examine the two objects in turn.

\subsection{PSR J1709$-$4429}
\label{subsec:j1709}
\begin{figure}
    \centering
    \includegraphics[width=\columnwidth]{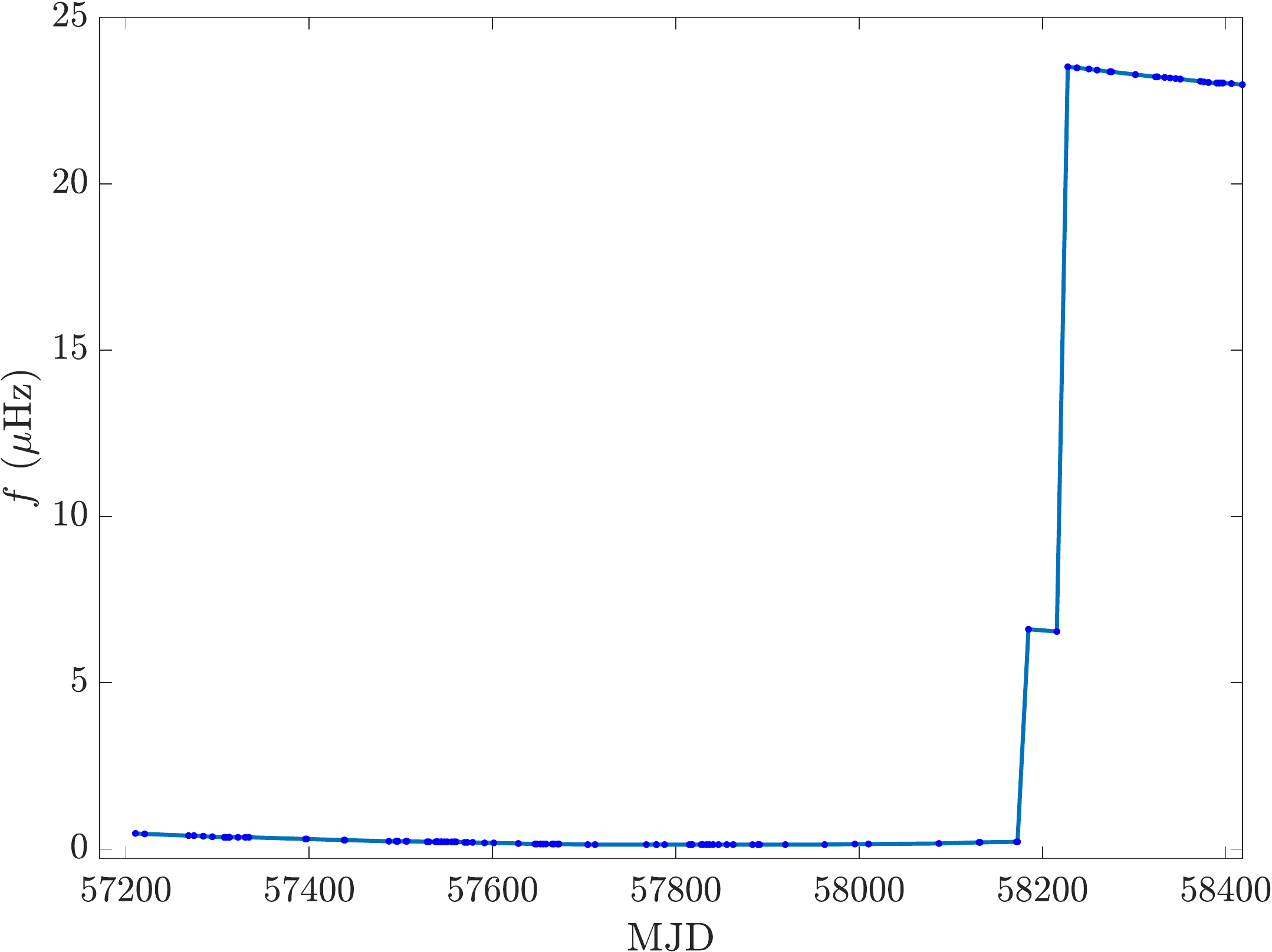}\\\vspace{0.4cm}
    \includegraphics[width=\columnwidth]{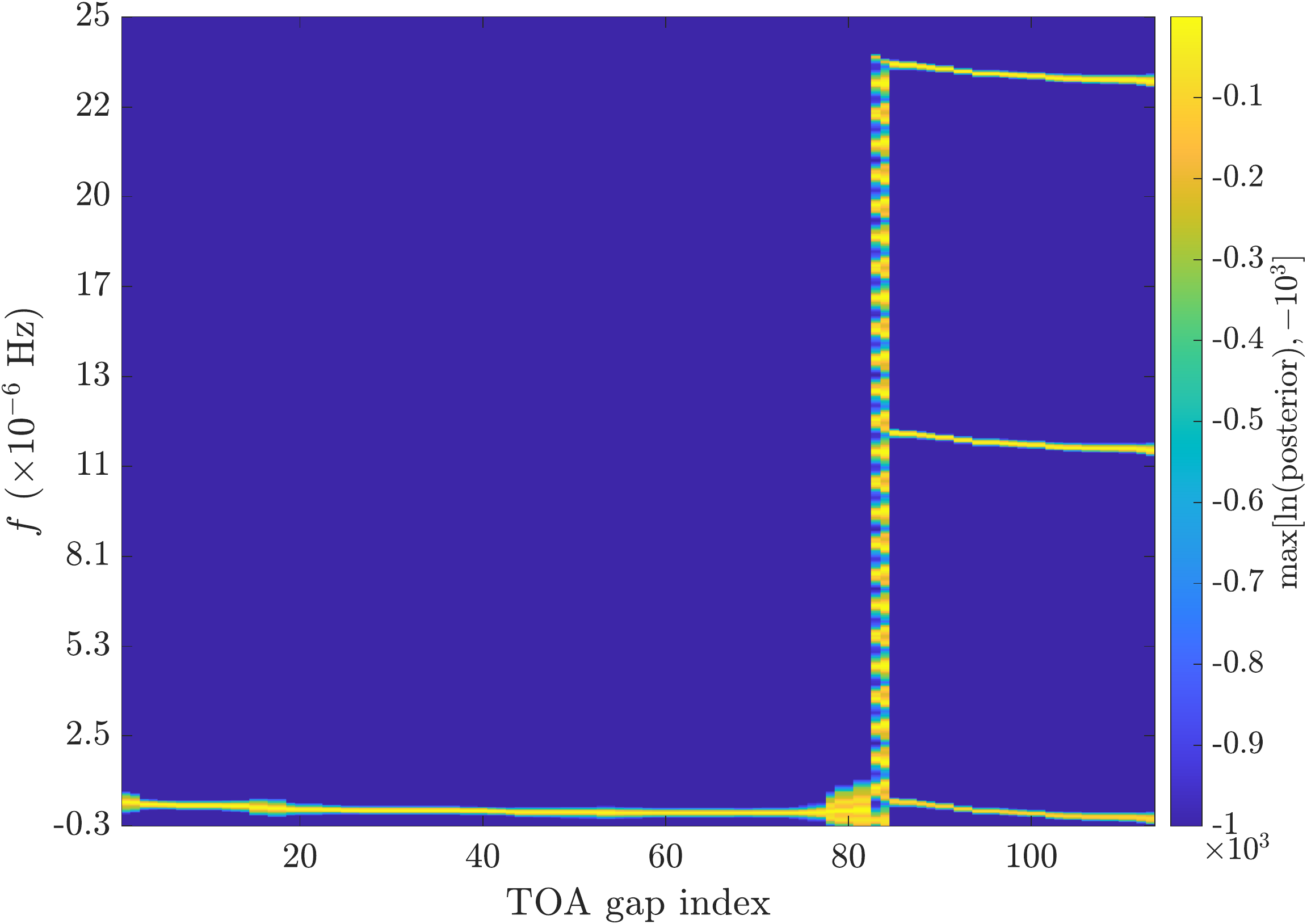}
    \caption{Recovered frequency path \emph{(top)} and posterior distribution of $f$ \emph{(bottom)} for the HMM analysis of the UTMOST observations of PSR J1709$-$4429, laid out as in Fig. \ref{fig:synth_J1452-6036_hmm}.
    Before the glitch, which occurs during the 81st ToA gap, the posterior distribution of $f$ is well-constrained, showing only a narrow band of support near $f = 0\,\mathrm{Hz}$.
    After the glitch, the posterior distribution of $f$ has support in three distinct $f$ regions, separated by $1/(1\,\text{sidereal day})$.
    The parameters used in this analysis are reported in Table \ref{tab:hmm_params_J1452-6036}.}
    \label{fig:J1709_hmm}
\end{figure}

\citet{lowerDetectionGlitchPulsar2018} reported a glitch in PSR J1709$-$4429 at MJD $58178 \pm 6$ of size $\Delta f/f = (52.4 \pm 0.1) \times 10^{-9}$, based on UTMOST data analysed with \textsc{tempo2} and \textsc{temponest}.
An updated parameter estimate was published by \citet{lowerUTMOSTPulsarTiming2020a}, who gave a glitch size of $\Delta f/f = (54.6 \pm 1.0) \times 10^{-9}$, again based only on UTMOST data.
An HMM-based analysis of these data recovers a glitch size of $\Delta f/f = 2405 \times 10^{-9}$, as shown in Fig. \ref{fig:J1709_hmm}.
The glitch size recovered by the HMM is larger than the \citet{lowerUTMOSTPulsarTiming2020a} result by $2.29 \times 10^{-5}\,\mathrm{Hz}$ -- roughly $2/(1\,\text{sidereal day})$.
This suggests that at least one of these methods is confounded by the periodicity of the observations, but without further information it is difficult to decide which glitch size is closer to the truth.

\begin{figure}
    \includegraphics[width=\columnwidth]{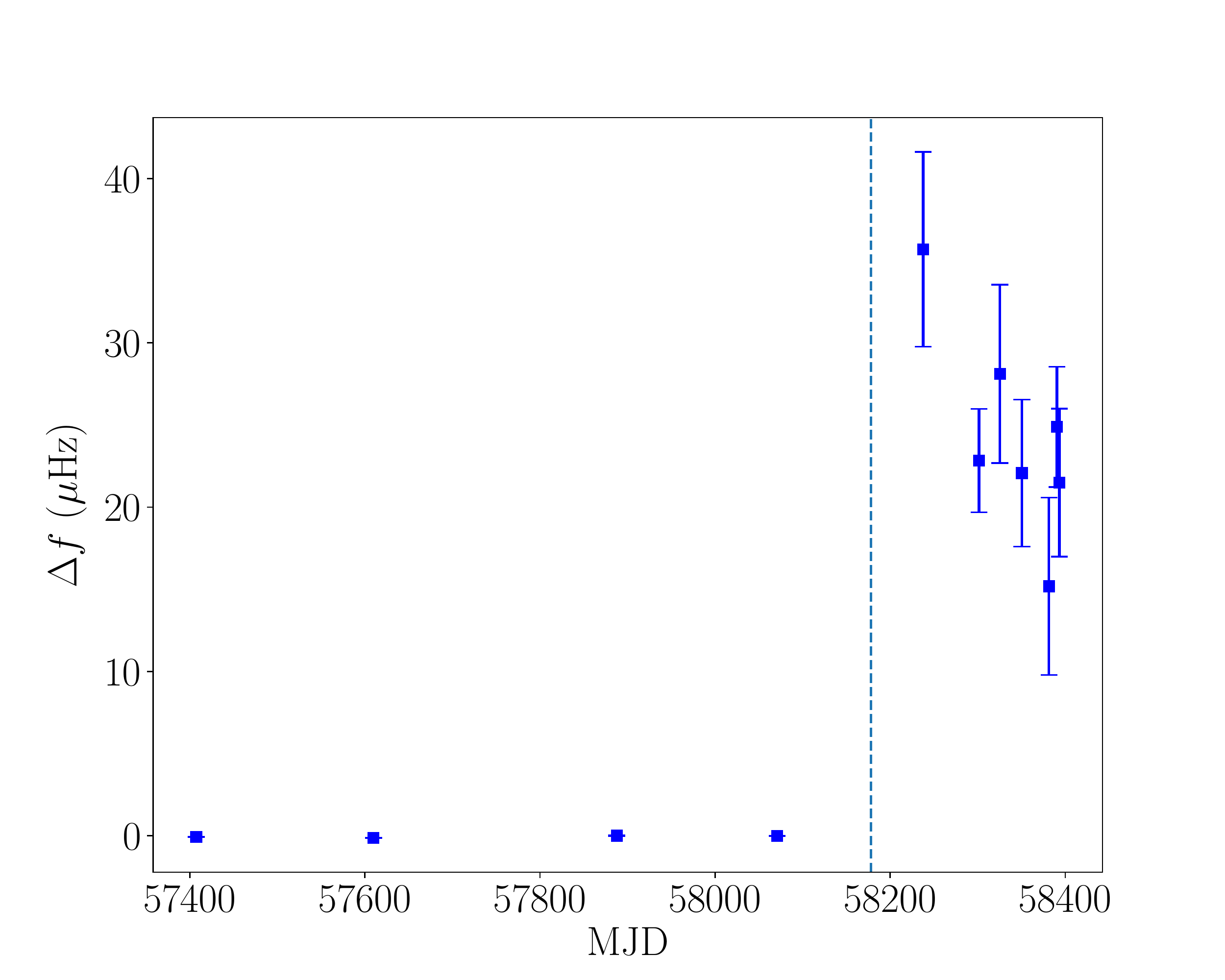}
    \caption{Local spin frequency estimation of the glitch size $\Delta f$ for the glitch in PSR J1709$-$4429 at MJD 58178 using UTMOST observations, laid out as in Fig. \ref{fig:local_f0}.
    The error bars on the pre-glitch frequency estimates are too small to be seen.}
    \label{fig:J1709-4429_2x}
\end{figure}

Re-processing the UTMOST data to produce multiple ToAs per observation session allows us to estimate the local spin frequency post-glitch, as described in Section \ref{subsubsec:local_f0}.
The results of this exercise are shown in Fig. \ref{fig:J1709-4429_2x}.
The post-glitch frequency measurements appear to be centred around roughly $2 \times 10^{-5}\,\mathrm{Hz}$, consistent with the HMM estimate.
Note that preceding the glitch, this pulsar was not observed with sufficiently long observations and sufficient sensitivity to extract enough high-quality ToAs per observing session for a useful local frequency fit. Each of the the pre-glitch frequency estimates therefore incorporate ToAs from observing sessions separated by multiple days, and the error bars are correspondingly much smaller than the post-glitch frequency estimates.
The use of widely separated ToAs is not a concern for the pre-glitch frequency estimates, as the pre-glitch frequency is not in question, having been well-measured by UTMOST before the switch to a periodic observing schedule, and this is the first glitch since the switch.

\begin{figure}
    \centering
    \includegraphics[width=\columnwidth]{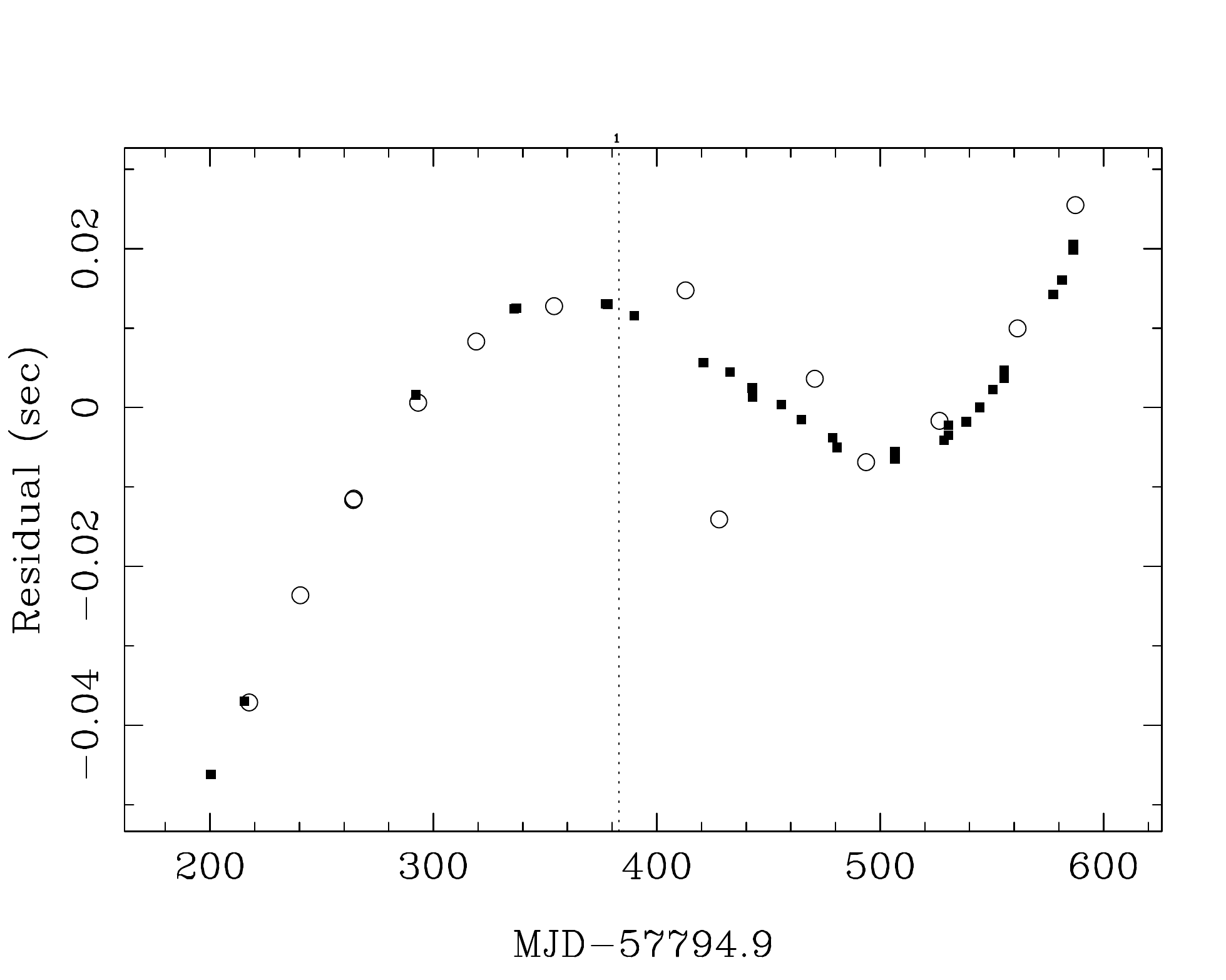}\\
    \includegraphics[width=\columnwidth]{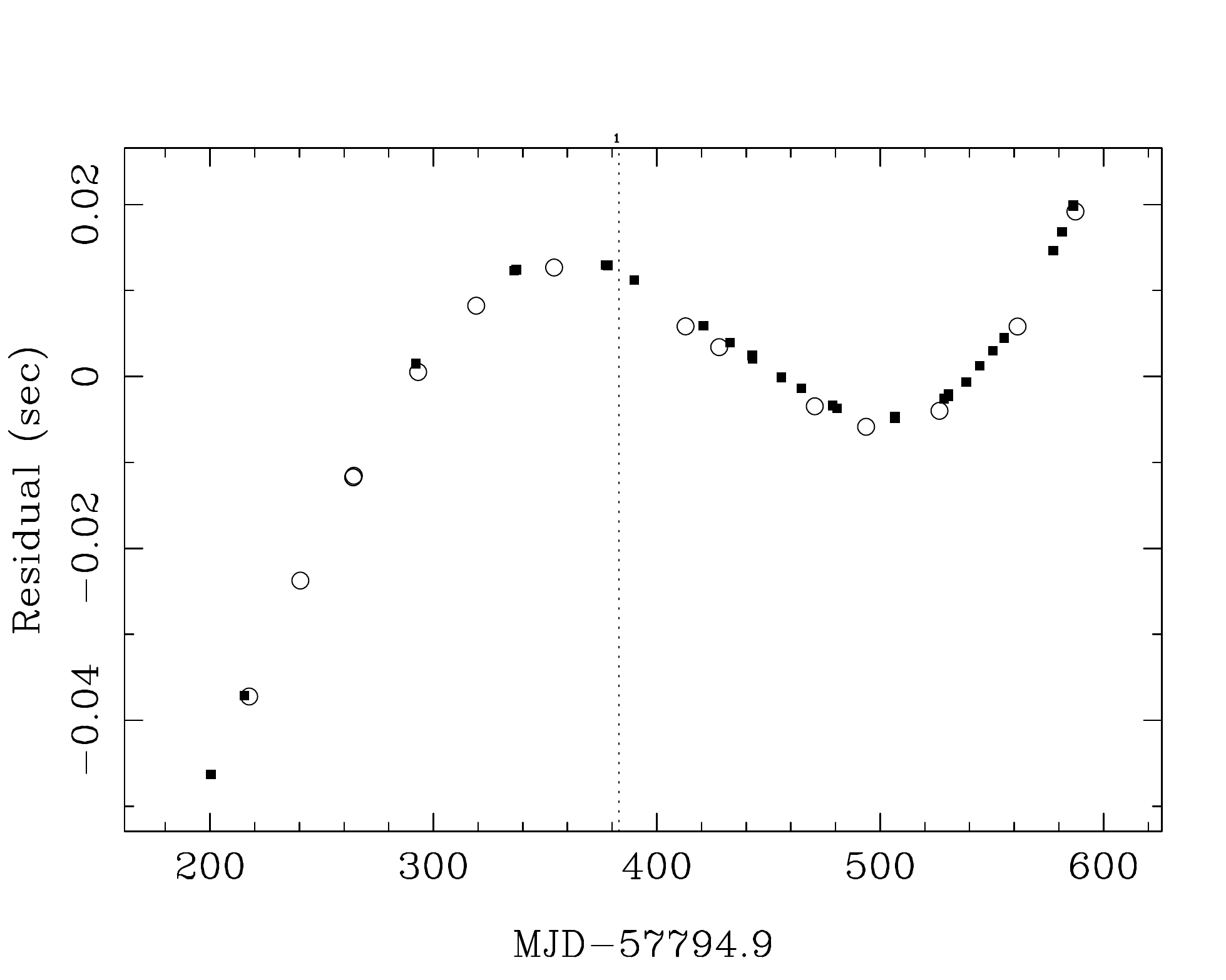}
    \caption{Timing residuals for combined UTMOST and Parkes observations of PSR J1709$-$4429 for two timing models with glitch sizes $\Delta f/f = 54.6 \times 10^{-9}$ \emph{(top)} and $\Delta f/f = 2429.7 \times 10^{-9}$ \emph{(bottom)}. In each plot the unphysical phase jump $\Delta \phi$ has also been adjusted to minimise the jump in residuals before and after the glitch. 
    UTMOST observations are shown as squares, and the Parkes observations are shown as circles.
    In both plots, all other timing model parameters besides $\Delta f$ and $\Delta \phi$ are identical to those published in the UTMOST data release.}
    \label{fig:J1709-4429}
\end{figure}

Fortunately, the pulsar timing programme carried out at the Parkes radio telescope has released public data covering the period during which the glitch was reported \citep{hobbsParkesObservatoryPulsar2011a}, and the Parkes pulsar timing programme does not schedule observations with the same regularity as UTMOST.
Hence, the combined UTMOST and Parkes data can be expected to estimate the glitch parameters better than the UTMOST data alone.
Fig. \ref{fig:J1709-4429} shows timing residuals for two glitch models with the combined UTMOST and Parkes data.
With the combined data, the glitch model with $\Delta f/f = 2429.7 \times 10^{-9}$ is clearly preferred, close to what was recovered in the HMM and local frequency estimation analyses.

As a final step, we re-estimate the glitch parameters with the combined UTMOST and Parkes data using \textsc{temponest} to fit for the timing model parameters and the noise parameters.
We start with a pulse numbering derived from a timing model with glitch size $\Delta f/f = 2429.7 \times 10^{-9}$.
The glitch model used follows \citet{lowerUTMOSTPulsarTiming2020a}, and is given by \begin{align} \Delta \phi_\text{g}(t) = \Delta \phi &+ \Delta f (t-t_\text{g}) \nonumber\\&+ \frac{1}{2}\Delta\dot{f} (t-t_\text{g})^2 - \Delta f_\text{d}\tau_\text{d} e^{-(t-t_\text{g})/\tau_\text{d}}.\label{eqn:j1709_glitch_model} \end{align}
The obtained glitch parameters are shown in Table \ref{tab:j1709_refit}.
The unphysical phase jump $\Delta\phi = -0.342 \pm 0.005$ arises because we choose to fix the glitch epoch at MJD $58178$, as in the analysis of \citet{lowerUTMOSTPulsarTiming2020a}.
We report only an upper limit on the size of the exponentially decaying term, $\Delta f_\text{d}$.
The posterior distribution for $\Delta f_\text{d}$ has support between the lower end of the prior range, $10^{-18}\,\mathrm{Hz}$, and $10^{-9}\,\mathrm{Hz}$, but no support above $10^{-9}\,\mathrm{Hz}$.
The decay timescale $\tau_d$ is unconstrained -- the posterior distribution has significant support across the entire prior range of $(1, 1000)\,\mathrm{d}$.
Previous glitches of this pulsar have been measured with an exponentially decaying component roughly 1\% as large as the permanent frequency jump, with a decay timescale of approximately $100\,\mathrm{d}$ \citep{yuDetection107Glitches2013}.

\begin{table*}
    \centering
    \caption{Estimated glitch parameters for the glitch in PSR J1709$-$4429 at MJD 58178. The glitch model parameters are defined in equation (\ref{eqn:j1709_glitch_model}).}
    \begin{tabular}{ccccccc}\hline
        & $t_\text{g}$ & $\Delta\phi$ & $(\Delta f + \Delta f_\text{d})/f$ & $\Delta \dot{f}/\dot{f}$ & $\Delta f_\text{d}/f$ & $\tau_\text{d}$ \\
        & MJD && $\times 10^{-9}$ & $\times 10^{-3}$ & $\times 10^{-9}$ & d \\\hline
        This work & $58178 \pm 6$ &  $-0.342 \pm 0.005$ & $2432.2 \pm 0.1$ & $4.7 \pm 0.3$ & $<0.1$ & --\\
        \citet{lowerUTMOSTPulsarTiming2020a} & $58178 \pm 6$& $0.372$ & $54.6 \pm 1.0$ & $1.06_{-0.43}^{+0.36}$ & $54.3 \pm 1.0$ & $99.1_{-9.6}^{+11.3}$ \\\hline
    \end{tabular}
    \label{tab:j1709_refit}
\end{table*}

\subsection{PSR J1452$-$6036}
\citet{lowerUTMOSTPulsarTiming2020a} also reported a glitch in PSR J1452$-$6036 at MJD $58600.29 \pm 0.05$, with a glitch size of $\Delta f/f = (270.7^{+0.3}_{-0.4}) \times 10^{-9}$ and no measured $\Delta\dot{f}$. In the following we keep $\Delta\dot{f}$ fixed at zero for simplicity.

Local frequency estimation with the UTMOST data does not constrain the glitch size -- the typical ToA uncertainty is high, roughly $1\,\mathrm{ms}$, and so the corresponding phase error is $\sim 6 \times 10^{-3}$.
Based on the arguments of Section \ref{subsubsec:local_f0}, an observation session long enough to break the degeneracy between glitch models would be roughly $1\,\mathrm{hr}$.
This is much longer than the actual $\sim 5\,\mathrm{min}$ observation sessions.
The maximum possible observation time in a single transit with UTMOST for this pulsar is somewhat longer, roughly 20 minutes, but not long enough to break the degeneracy.
In general the maximum time that UTMOST can observe a given pulsar depends strongly on its declination: pulsars near the ecliptic transit the primary beam in approximately 10 minutes, while pulsars near the south celestial pole remain in the primary beam for hours.
However, most pulsars are not routinely observed by UTMOST for more than 10 minutes at a time \citep{jankowskiUTMOSTPulsarTiming2019}.
Alternatively, the per-ToA uncertainty which would allow a $5\,\mathrm{min}$ observation session to break the degeneracy is roughly $0.1\,\mathrm{ms}$.

\begin{figure}
    \includegraphics[width=\columnwidth]{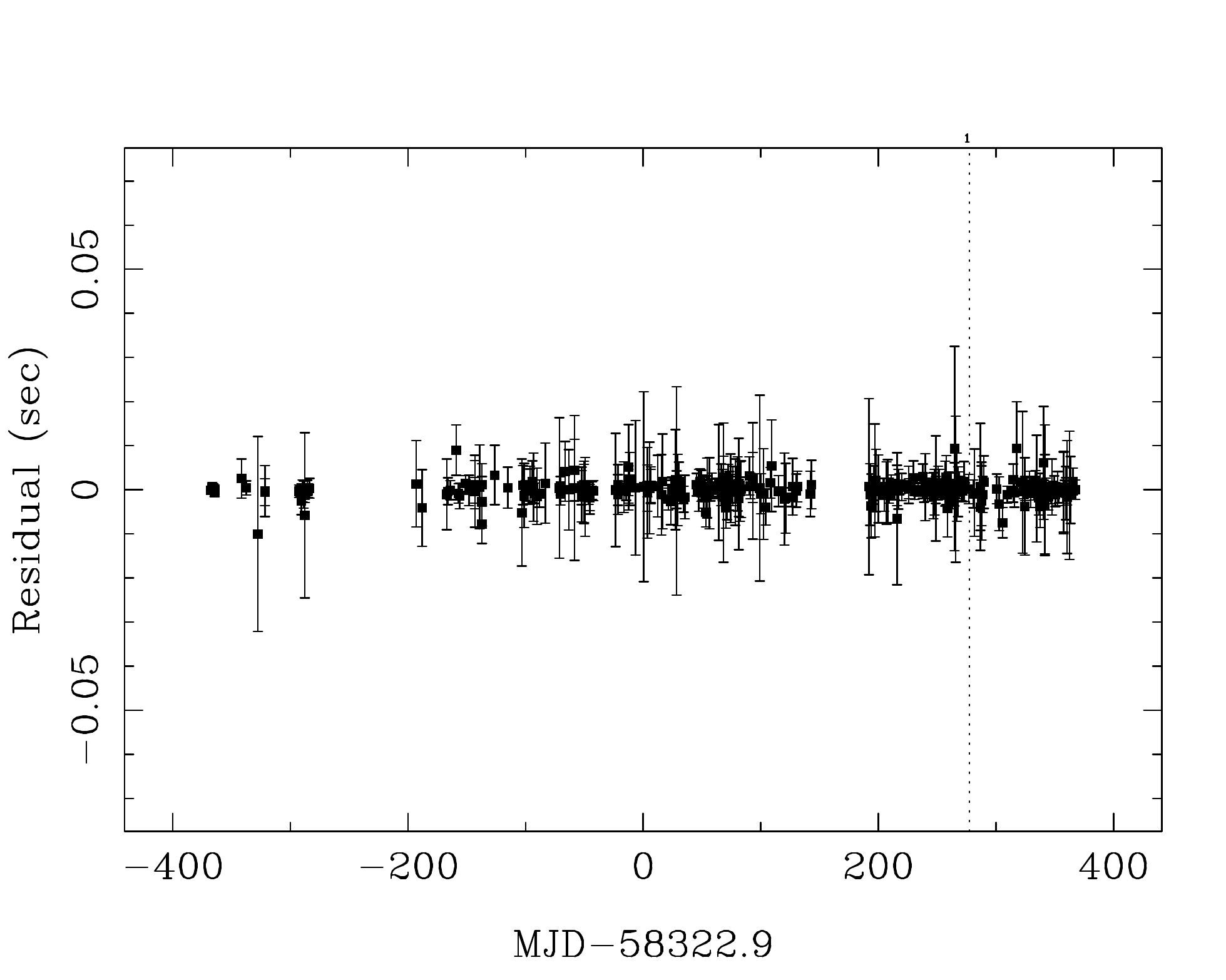}\\
    \includegraphics[width=\columnwidth]{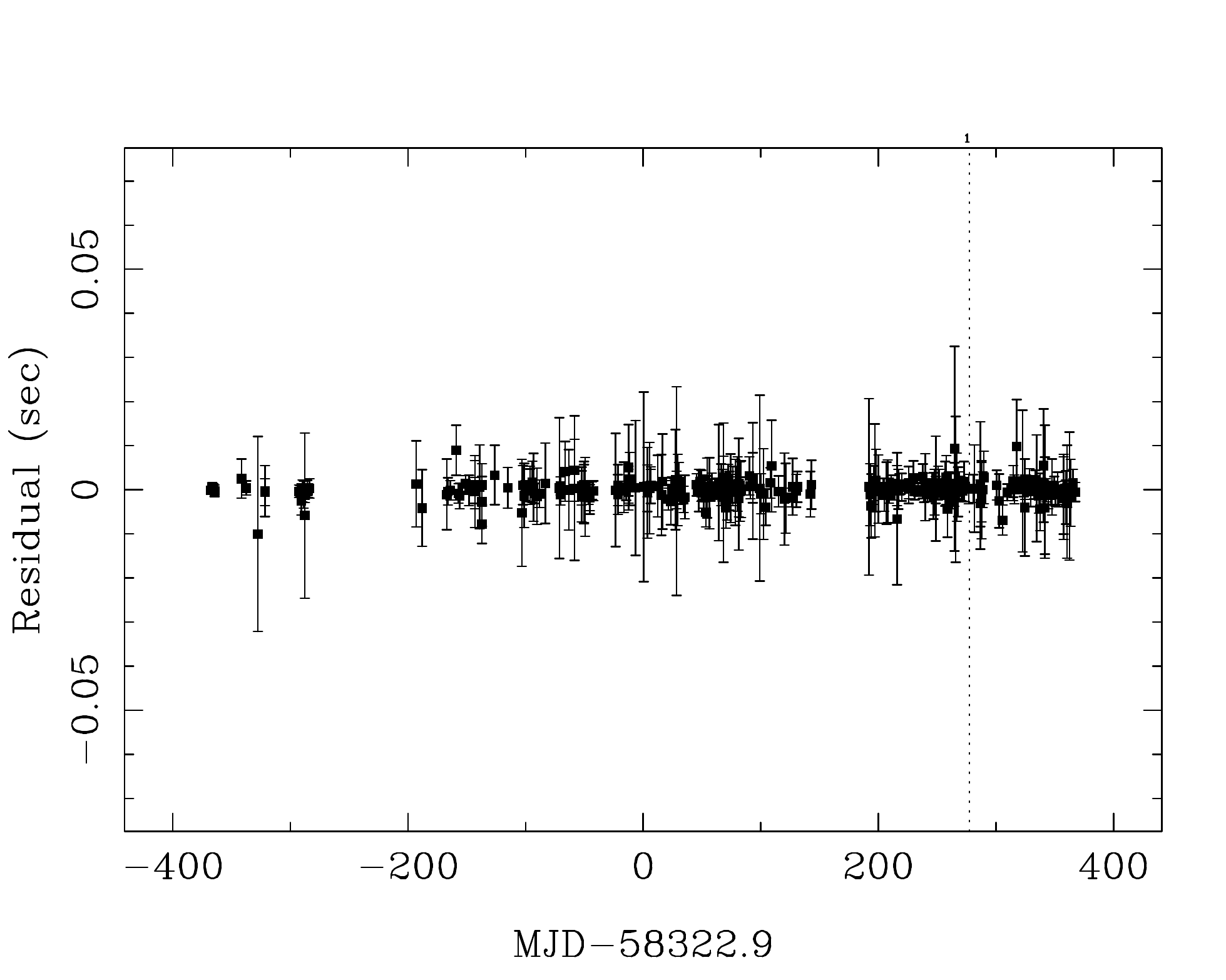}\\
    \includegraphics[width=\columnwidth]{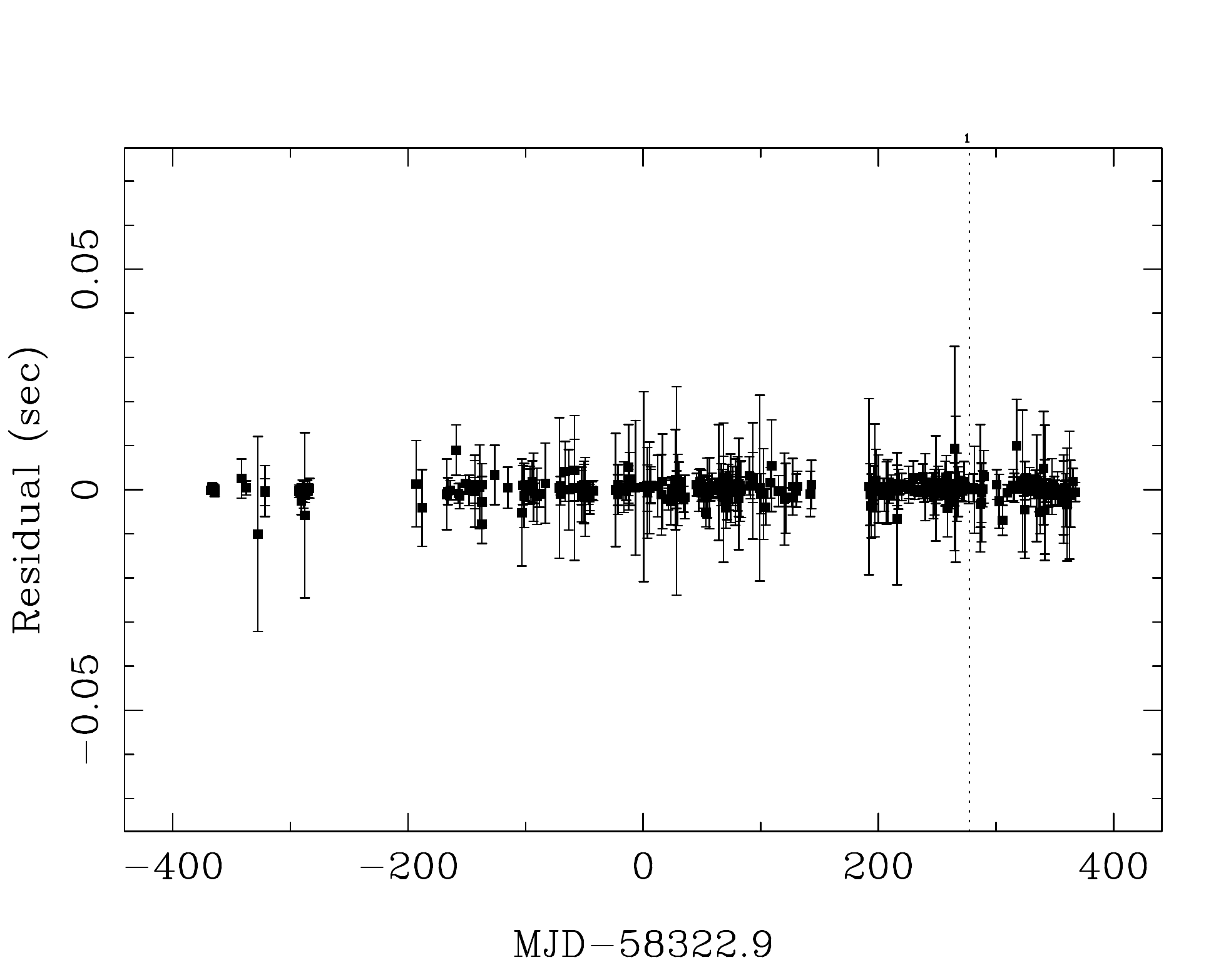}
    \caption{Timing residuals for UTMOST observations of PSR J1452$-$6036.
    Glitches are indicated by the vertical dotted line, with sizes $\Delta f/f = 270 \times 10^{-9}$ \emph{(top)}, $2069 \times 10^{-9}$ \emph{(middle)}, and $3868 \times 10^{-9}$ \emph{(bottom)}.
    The post-glitch residuals appear to be identical, despite the significant differences in glitch sizes in the three timing models.
    Other timing model parameters match those released in the UTMOST data release.}   
     \label{fig:J1452-6036_glitches}
\end{figure}

We may apply the arguments of Section \ref{subsec:detecting} to search for excess post-glitch phase residuals due to misestimation of the glitch size.
We set $P_\text{fa} = 0.01$, and find that the threshold calculated according to equation (\ref{eqn:error_detection_thresh}) is $\gamma_\mathrm{th} = 1.99$.
The probability of detection given by equation (\ref{eqn:error_detection_PD}) is only $P_\mathrm{d} = 0.07$.
We search for the induced residual signal defined by equation (\ref{eqn:error:detection_signal}) in the residuals of three glitch models: the originally reported model, a model with $\Delta f$ increased by $1/(1\,\text{sidereal day})$ to give $\Delta f/f = 2069 \times 10^{-9}$, and a model with $\Delta f$ increased by $2/(1\,\text{sidereal day})$ to give $\Delta f/f = 3868 \times 10^{-9}$.
The residuals for each glitch model are shown in Fig. \ref{fig:J1452-6036_glitches}.
We find in each case that the test statistic $\gamma = \mathbfit{R}^\mathrm{T}\mathbfss{C}^{-1}\mathbfit{S}$ does not exceed $\gamma_\mathrm{th}$, i.e. we do not detect an induced phase error.
$\gamma$ is largest for the glitch model with $\Delta f/f = 3868 \times 10^{-9}$, where we calculate $\gamma = 0.62$.
As $P_\mathrm{d}$ is low, this non-detection does not allow us to constrain the true parameters of the glitch.

\citet{jankowskiConstraintsWidebandRadiative2021} recently reported on wideband observations of this pulsar at the Parkes radio telescope which by chance happen to lie on either side of this glitch.
Combining these observations with the UTMOST data, they report a glitch size $\Delta f/f = 270.52(3) \times 10^{-9}$, consistent with the value found by \citet{lowerUTMOSTPulsarTiming2020a}.
As with the PSR J1709$-$4429 observations, the Parkes observations of PSR J1452$-$6036 are not on the same schedule as the UTMOST observations, and so the estimate of $\Delta f$ from the combined data is not confounded by the periodicity of the UTMOST observations.

\section{Conclusion}
In this paper we show that periodic scheduling of pulsar observations can lead to erroneous estimates of frequency in pulsar timing models if the frequency is not well-measured \emph{a priori}. We examine in detail the effect this has on the estimation of pulsar glitch parameters.
Specifically, the estimated permanent frequency jump $\Delta f$ may be displaced from its true value by an integer multiple of $1/T$ where $T$ is the scheduling period.
We find that in certain existing datasets the excess timing residuals induced by misestimating the spin frequency of the pulsar due to periodic scheduling are comparable to the stochastic residuals induced by ToA measurement error.

We find that ``by-eye'' attempts to restore phase connection through the use of pulse numbering can bias the recovered glitch size towards smaller values.
When the true value of $\Delta f$ is larger than $1/T$, this bias yields incorrect estimates of $\Delta f$.
Local frequency estimation can mitigate this bias, as long as the ToAs used in the fits are sufficiently accurate.
An HMM-based approach may also fail to recover the correct $\Delta f$.
However, the HMM does not appear to be biased in the same way as the phase-coherent timing-based method, and the existence of multiple solutions is readily apparent from a brief inspection of the products of the analysis, e.g. the posteriors of $f(t_1), \ldots, f(t_N)$.

We re-evaluate two glitches detected by UTMOST, in PSR J1709$-$4429 and PSR J1452$-$6036.
In the case of PSR J1709$-$4429 additional public data from the Parkes radio telescope breaks the degeneracy in glitch models, and we recover a new glitch size $\Delta f/f = (2432.2 \pm 0.1)\times 10^{-9}$, much larger than the previously reported value $\Delta f/f = (54.6 \pm 1.0) \times 10^{-9}$.
For PSR J1452$-$6036, we attempt to detect the presence of phase residuals induced by an incorrect choice of glitch model by cross-correlating the observed residuals with the expected induced signal.
However we are unable to definitively confirm or exclude the previously reported glitch model using UTMOST data alone, because the noise is relatively high, and there are relatively few post-glitch observations.
Recently released complementary observations from the Parkes radio telescope break the degeneracy and confirm that the glitch size is $\Delta f/f = 270.52(3) \times 10^{-9}$ \citep{jankowskiConstraintsWidebandRadiative2021}.

In view of the issues raised here, we recommend that wherever possible, future observing campaigns aimed at glitch measurement should avoid periodic observation scheduling.
In cases where periodic observations are unavoidable, such as the CHIME/Pulsar \citep{ngPulsarScienceCHIME2018, goodFirstDiscoveryNew2020} and future UTMOST-2D \citep{venkatramankrishnanUTMOSTSurveyMagnetars2020} programmes, care should be taken when making inferences about the frequency of a pulsar, particularly after a glitch has occurred.
A small number of complementary observations from another observatory can also help break the degeneracy between glitch models.

\section*{Acknowledgements}
Parts of this research are supported by the Australian Research Council (ARC) Centre of Excellence for Gravitational Wave Discovery (OzGrav) (project number CE170100004) and ARC Discovery Project DP170103625.
L. Dunn is supported by an Australian Government Research Training Program Scholarship and by the Rowden White Scholarship.
M. E. Lower receives support from the ARC Laureate Fellowship FL15010014 and CSIRO Astronomy and Space Science.
The authors are grateful to Bill Moran, Rob Evans, and especially Sofia Suvorova for work on the hidden Markov model and useful discussions.
The authors are also grateful to the anonymous referee for their helpful comments.

\section*{Data availability}
{The public UTMOST data underlying} this work are available at \url{https://github.com/M    olonglo/TimingDataRelease1/}.
The other data underlying this work will be shared on reasonable request to the corresponding author.

\bibliography{main}{}

\begin{thebibliography}{}
\makeatletter
\relax
\def\mn@urlcharsother{\let\do\@makeother \do\$\do\&\do\#\do\^\do\_\do\%\do\~}
\def\mn@doi{\begingroup\mn@urlcharsother \@ifnextchar [ {\mn@doi@}
  {\mn@doi@[]}}
\def\mn@doi@[#1]#2{\def\@tempa{#1}\ifx\@tempa\@empty \href
  {http://dx.doi.org/#2} {doi:#2}\else \href {http://dx.doi.org/#2} {#1}\fi
  \endgroup}
\def\mn@eprint#1#2{\mn@eprint@#1:#2::\@nil}
\def\mn@eprint@arXiv#1{\href {http://arxiv.org/abs/#1} {{\tt arXiv:#1}}}
\def\mn@eprint@dblp#1{\href {http://dblp.uni-trier.de/rec/bibtex/#1.xml}
  {dblp:#1}}
\def\mn@eprint@#1:#2:#3:#4\@nil{\def\@tempa {#1}\def\@tempb {#2}\def\@tempc
  {#3}\ifx \@tempc \@empty \let \@tempc \@tempb \let \@tempb \@tempa \fi \ifx
  \@tempb \@empty \def\@tempb {arXiv}\fi \@ifundefined
  {mn@eprint@\@tempb}{\@tempb:\@tempc}{\expandafter \expandafter \csname
  mn@eprint@\@tempb\endcsname \expandafter{\@tempc}}}

\bibitem[\protect\citeauthoryear{Ashton, Lasky, Graber  \& Palfreyman}{Ashton
  et~al.}{2019}]{ashtonRotationalEvolutionVela2019}
Ashton G.,  Lasky P.~D.,  Graber V.,   Palfreyman J.,  2019, \mn@doi [Nature
  Astronomy] {10.1038/s41550-019-0844-6}

\bibitem[\protect\citeauthoryear{Bailes et~al.,}{Bailes
  et~al.}{2017}]{bailesUTMOSTHybridDigital2017}
Bailes M.,  et~al., 2017, \mn@doi [Publications of the Astronomical Society of
  Australia] {10.1017/pasa.2017.39}

\bibitem[\protect\citeauthoryear{Basu, Joshi, Krishnakumar, Bhattacharya,
  Nandi, Bandhopadhay, Char  \& Manoharan}{Basu
  et~al.}{2020}]{basuObservedGlitchesEight2020}
Basu A.,  Joshi B.~C.,  Krishnakumar M.~A.,  Bhattacharya D.,  Nandi R.,
  Bandhopadhay D.,  Char P.,   Manoharan P.~K.,  2020, \mn@doi [Monthly Notices
  of the Royal Astronomical Society] {10.1093/mnras/stz3230}, 491, 3182

\bibitem[\protect\citeauthoryear{Dodson, McCulloch  \& Lewis}{Dodson
  et~al.}{2002}]{dodsonHighTimeResolution2002}
Dodson R.~G.,  McCulloch P.~M.,   Lewis D.~R.,  2002, \mn@doi [The
  Astrophysical Journal Letters] {10.1086/339068}, 564, L85

\bibitem[\protect\citeauthoryear{Edwards, Hobbs  \& Manchester}{Edwards
  et~al.}{2006}]{edwardsTEMPO2NewPulsar2006}
Edwards R.~T.,  Hobbs G.~B.,   Manchester R.~N.,  2006, \mn@doi [Monthly
  Notices of the Royal Astronomical Society]
  {10.1111/j.1365-2966.2006.10870.x}, 372, 1549

\bibitem[\protect\citeauthoryear{Espinoza, Lyne, Stappers  \& Kramer}{Espinoza
  et~al.}{2011}]{espinozaStudy315Glitches2011}
Espinoza C.~M.,  Lyne A.~G.,  Stappers B.~W.,   Kramer M.,  2011, \mn@doi
  [Monthly Notices of the Royal Astronomical Society]
  {10.1111/j.1365-2966.2011.18503.x}, 414, 1679

\bibitem[\protect\citeauthoryear{Espinoza, Antonopoulou, Stappers, Watts  \&
  Lyne}{Espinoza et~al.}{2014}]{espinozaNeutronStarGlitches2014}
Espinoza C.~M.,  Antonopoulou D.,  Stappers B.~W.,  Watts A.,   Lyne A.~G.,
  2014, \mn@doi [Monthly Notices of the Royal Astronomical Society]
  {10.1093/mnras/stu395}, 440, 2755

\bibitem[\protect\citeauthoryear{Feroz, Hobson  \& Bridges}{Feroz
  et~al.}{2009}]{ferozMultiNestEfficientRobust2009}
Feroz F.,  Hobson M.~P.,   Bridges M.,  2009, \mn@doi [Monthly Notices of the
  Royal Astronomical Society] {10.1111/j.1365-2966.2009.14548.x}, 398, 1601

\bibitem[\protect\citeauthoryear{Good et~al.,}{Good
  et~al.}{2020}]{goodFirstDiscoveryNew2020}
Good D.~C.,  et~al., 2020, arXiv, 2012.02320

\bibitem[\protect\citeauthoryear{Hobbs, Edwards  \& Manchester}{Hobbs
  et~al.}{2006}]{hobbsTEMPO2NewPulsartiming2006}
Hobbs G.~B.,  Edwards R.~T.,   Manchester R.~N.,  2006, \mn@doi [Monthly
  Notices of the Royal Astronomical Society]
  {10.1111/j.1365-2966.2006.10302.x}, 369, 655

\bibitem[\protect\citeauthoryear{Hobbs et~al.,}{Hobbs
  et~al.}{2011}]{hobbsParkesObservatoryPulsar2011a}
Hobbs G.,  et~al., 2011, \mn@doi [Publications of the Astronomical Society of
  Australia] {10.1071/AS11016}, 28, 202

\bibitem[\protect\citeauthoryear{Jankowski et~al.,}{Jankowski
  et~al.}{2019}]{jankowskiUTMOSTPulsarTiming2019}
Jankowski F.,  et~al., 2019, \mn@doi [Monthly Notices of the Royal Astronomical
  Society] {10.1093/mnras/sty3390}, 484, 3691

\bibitem[\protect\citeauthoryear{Jankowski, Keane  \& Stappers}{Jankowski
  et~al.}{2021}]{jankowskiConstraintsWidebandRadiative2021}
Jankowski F.,  Keane E.~F.,   Stappers B.~W.,  2021, arXiv:2103.09869
  [astro-ph]

\bibitem[\protect\citeauthoryear{Janssen \& Stappers}{Janssen \&
  Stappers}{2006}]{janssen30GlitchesSlow2006}
Janssen G.~H.,  Stappers B.~W.,  2006, \mn@doi [Astronomy \& Astrophysics]
  {10.1051/0004-6361:20065267}, 457, 611

\bibitem[\protect\citeauthoryear{Lentati, Alexander, Hobson, Feroz, {van
  Haasteren}, Lee  \& Shannon}{Lentati
  et~al.}{2014}]{lentatiTemponestBayesianApproach2014}
Lentati L.,  Alexander P.,  Hobson M.~P.,  Feroz F.,  {van Haasteren} R.,  Lee
  K.~J.,   Shannon R.~M.,  2014, \mn@doi [Monthly Notices of the Royal
  Astronomical Society] {10.1093/mnras/stt2122}, 437, 3004

\bibitem[\protect\citeauthoryear{Levy}{Levy}{2008}]{levyPrinciplesSignalDetection2008}
Levy B.~C.,  2008, Principles of {{Signal Detection}} and {{Parameter
  Estimation}}.
{Springer US}, \mn@doi{10.1007/978-0-387-76544-0}

\bibitem[\protect\citeauthoryear{Lorimer \& Kramer}{Lorimer \&
  Kramer}{2004}]{lorimerHandbookPulsarAstronomy2004}
Lorimer D.~R.,  Kramer M.,  2004, Handbook of {{Pulsar Astronomy}}.
{Cambridge University Press}

\bibitem[\protect\citeauthoryear{Lower et~al.,}{Lower
  et~al.}{2018}]{lowerDetectionGlitchPulsar2018}
Lower M.~E.,  et~al., 2018, \mn@doi [Research Notes of the AAS]
  {10.3847/2515-5172/aad7bc}, 2, 139

\bibitem[\protect\citeauthoryear{Lower et~al.,}{Lower
  et~al.}{2020}]{lowerUTMOSTPulsarTiming2020a}
Lower M.~E.,  et~al., 2020, \mn@doi [Monthly Notices of the Royal Astronomical
  Society] {10.1093/mnras/staa615}, 494, 228

\bibitem[\protect\citeauthoryear{Marshall, Gotthelf, Middleditch, Wang  \&
  Zhang}{Marshall et~al.}{2004}]{marshallBigGlitcherRotation2004}
Marshall F.~E.,  Gotthelf E.~V.,  Middleditch J.,  Wang Q.~D.,   Zhang W.,
  2004, \mn@doi [The Astrophysical Journal] {10.1086/381567}, 603, 682

\bibitem[\protect\citeauthoryear{Melatos, Dunn, Suvorova, Moran  \&
  Evans}{Melatos et~al.}{2020}]{melatosPulsarGlitchDetection2020}
Melatos A.,  Dunn L.~M.,  Suvorova S.,  Moran W.,   Evans R.~J.,  2020, \mn@doi
  [The Astrophysical Journal] {10.3847/1538-4357/ab9178}, 896, 78

\bibitem[\protect\citeauthoryear{Ng}{Ng}{2018}]{ngPulsarScienceCHIME2018}
Ng C.,  2018, \mn@doi [Proceedings of the International Astronomical Union]
  {10.1017/S1743921317010638}, 337, 179

\bibitem[\protect\citeauthoryear{Parthasarathy et~al.,}{Parthasarathy
  et~al.}{2020}]{parthasarathyTimingYoungRadio2020}
Parthasarathy A.,  et~al., 2020, \mn@doi [Monthly Notices of the Royal
  Astronomical Society] {10.1093/mnras/staa882}, 494, 2012

\bibitem[\protect\citeauthoryear{Rabiner}{Rabiner}{1989}]{rabinerTutorialHiddenMarkov1989}
Rabiner L.~R.,  1989, Proceedings of the IEEE, 77, 257

\bibitem[\protect\citeauthoryear{Shannon, Lentati, Kerr, Johnston, Hobbs  \&
  Manchester}{Shannon
  et~al.}{2016}]{shannonCharacterizingRotationalIrregularities2016}
Shannon R.~M.,  Lentati L.~T.,  Kerr M.,  Johnston S.,  Hobbs G.,   Manchester
  R.~N.,  2016, \mn@doi [Monthly Notices of the Royal Astronomical Society]
  {10.1093/mnras/stw842}, 459, 3104

\bibitem[\protect\citeauthoryear{Taylor}{Taylor}{1992}]{taylorPulsarTimingRelativistic1992}
Taylor J.~H.,  1992, \mn@doi [Philosophical Transactions of the Royal Society
  of London. Series A: Physical and Engineering Sciences]
  {10.1098/rsta.1992.0088}, 341, 117

\bibitem[\protect\citeauthoryear{Venkatraman~Krishnan
  et~al.,}{Venkatraman~Krishnan
  et~al.}{2020}]{venkatramankrishnanUTMOSTSurveyMagnetars2020}
Venkatraman~Krishnan V.,  et~al., 2020, \mn@doi [Monthly Notices of the Royal
  Astronomical Society] {10.1093/mnras/staa111}, 492, 4752

\bibitem[\protect\citeauthoryear{Wong, Backer  \& Lyne}{Wong
  et~al.}{2001}]{wongObservationsSeriesSix2001}
Wong T.,  Backer D.~C.,   Lyne A.~G.,  2001, \mn@doi [The Astrophysical
  Journal] {10.1086/318657}, 548, 447

\bibitem[\protect\citeauthoryear{Yu \& Liu}{Yu \&
  Liu}{2017}]{yuDetectionProbabilityNeutron2017}
Yu M.,  Liu Q.,  2017, \mn@doi [Monthly Notices of the Royal Astronomical
  Society] {10.1093/mnras/stx702}, 3041, 3031

\bibitem[\protect\citeauthoryear{Yu et~al.,}{Yu
  et~al.}{2013}]{yuDetection107Glitches2013}
Yu M.,  et~al., 2013, \mn@doi [Monthly Notices of the Royal Astronomical
  Society] {10.1093/mnras/sts366}, 429, 688

\makeatother
\end{thebibliography}
\bibliographystyle{mnras}

\appendix{
\section{HMM recipe and parameters}
\label{apdx:hmm_params}
The HMM-based analysis described in Section \ref{subsec:hmm} involves choosing a number of input parameters.
In this appendix we briefly describe these choices.
For a more detailed discussion of the considerations involved in choosing the analysis parameters, see \citet{melatosPulsarGlitchDetection2020}.
There are broadly three classes of parameters involved: those which specify the $(f, \dot{f})$ pairs under consideration, those which specify the connection between observations (ToAs) and $(f, \dot{f})$ pairs, and those which specify the the probabilities of transitions between $(f, \dot{f})$ pairs.
A complete list of parameters is given in Table \ref{tab:hmm_params_J1452-6036}.

We specify the allowed range of frequencies and frequency derivatives in two stages.
First, we specify a fiducial phase evolution by fixing the frequency $f_0$ and frequency derivative $\dot{f}_0$ at a reference epoch $T_0$, derived from a Taylor expansion computed by \textsc{tempo2}.
The HMM tracks deviations away from this fiducial model on a discrete grid in the $f$-$\dot{f}$ plane.
The range of allowed deviations is specified by lower ($f_-$, $\dot{f}_-$) and upper ($f_+$, $\dot{f}_+$) bounds.
The discretization is specified by bin sizes $\eta_f$ and $\eta_{\dot{f}}$.

To incorporate timing noise, we adopt a simple prescription which drives the second frequency derivative with a white noise term $\xi(t)$ satisfying $\langle\xi(t)\xi(t')\rangle = \sigma^2\delta(t-t')$ [see Section 3.4 in \citet{melatosPulsarGlitchDetection2020}].
Other, similar presciptions yield similar results \citep{melatosPulsarGlitchDetection2020}.
The free parameter $\sigma$ controls the magnitude of the timing noise in the model.
The discrete nature of the $f$-$\dot{f}$ grid sets a lower bound on $\sigma$: errors in the estimation of $\dot{f}$, caused by the finite bin width $\eta_{\dot{f}}$, cause the frequency to spuriously wander across a ToA gap by an amount $\delta f_\text{bin} \sim \eta_{\dot{f}}\Delta t_i$.
It is desirable that the HMM ``correct'' this spurious wandering, through the freedom allowed in the timing noise model.
With $\xi(t)$ defined as above, the frequency wandering across $\Delta t_i$ is given by $\delta f_\text{TN} = \sigma (\Delta t_i)^{3/2}$.
Equating $\delta f_\text{bin}$ and $\delta f_\text{TN}$ with $\Delta t_i$ replaced by its average over all ToAs, $\langle \Delta t_i\rangle$, suggests the following approximate lower bound, \begin{equation} \sigma \geq \eta_{\dot{f}}\langle \Delta t_i \rangle^{-1/2}. \end{equation}
In the analyses presented here, we choose $\sigma = \eta_{\dot{f}}\langle \Delta t_i \rangle^{-1/2}$.

Given a particular ToA gap $\Delta t_i$, the probability of a particular $(f, \dot{f})$ state is calculated as a von Mises distribution \citep{melatosPulsarGlitchDetection2020} \begin{equation} L(\Delta t_i; f, \dot{f}) = [2\pi I_0(\kappa)]^{-1}\exp\{\kappa\cos[2\pi(\Delta t_i f - \Delta t_i^2\dot{f}/2 + \Delta \Phi_i)]\}, \label{eqn:von_mises} \end{equation}
where $I_0(x)$ is the modified Bessel function of the first kind, $\kappa$ is a free parameter known as the concentration, and $\Delta\Phi_i$ is the phase contribution over the ToA gap from the fiducial phase model specified by $f_0$, $\dot{f}_0$, and the reference epoch $T_0$.
This distribution is peaked when the number of cycles accumulated across the ToA gap, $\Delta t_i f - \Delta t_i^2\dot{f}/2 + \Delta\Phi_i$, is an integer (the minus sign appears because we employ a backwards Taylor expansion).
To a good approximation we may identify $\kappa$ as the reciprocal of the squared uncertainty of the phase.
There are two important contributions to $\kappa$.
One is the uncertainty on individual ToAs.
For each ToA gap, the phase uncertainties on the two ToAs which bracket the gap, $\sigma_\text{ToA,1}$ and $\sigma_\text{ToA,2}$, contribute independently to the total phase uncertainty across the gap.
The second contribution comes from the discrete $f$-$\dot{f}$ grid.
The binning in frequency contributes a phase uncertainty $\eta_f \Delta t_i$, and the binning in frequency derivative contributes a phase uncertainty $\eta_{\dot{f}}\Delta t_i^2/2$.
Combining all of these uncertainties in quadrature, and recalling the identification of $\kappa$ with the reciprocal squared phase uncertainty, we arrive at the expression for $\kappa$ used in these analyses: \begin{equation} \kappa = \left[\sigma_\text{ToA,1}^2 + \sigma_\text{ToA,2}^2 + (\eta_f \Delta t_i)^2 + (\eta_{\dot{f}}\Delta t_i^2/2)^2\right]^{-1}. \label{eqn:kappa}\end{equation}
Note that $\kappa$ depends strongly on the length of the ToA gap, and therefore is recalculated for each gap in the dataset.
More details on the theoretical underpinnings of (\ref{eqn:von_mises}) and (\ref{eqn:kappa}) appear in Section 3.3 and appendix C in \citet{melatosPulsarGlitchDetection2020}.

Finally, we choose a Bayes factor threshold to be used in model selection via the greedy hierarchical algorithm described in Section 4.2 of \citet{melatosPulsarGlitchDetection2020}, when models containing glitches are compared against models with no or fewer glitches.
Informed by synthetic data tests described by \citet{melatosPulsarGlitchDetection2020}, we choose here a threshold of $K_\mathrm{th} = 10^{1/2}$. 
\begin{table*}
    \centering
    \caption{HMM parameters for the analysis (described in Section \ref{subsec:hmm}) of the synthetic dataset described in Section \ref{subsec:glitch_worked_example} and the PSR J1709$-$4429 analysis described in Section \ref{subsec:j1709}. Parameters marked with an asterisk are from the UTMOST data release.}
    \begin{tabular}{lllll}
        \hline
        Parameter & Symbol & Units & Synthetic data & \newcommand{\pointRaised}[2]{\medskip \hrule \noindent 
        \textsl{{\fontseries{b}\selectfont #1}: #2}} 
\newcommand{\reply}{\noindent \textbf{Reply}:\ } PSR J1709$-$4429 \\\hline
        Timing model reference epoch* & $T_0$ & MJD & $57600$ & $57600$ \\
        Fiducial frequency* & $f_0$ & Hz & $6.45193972751$ & $9.75429004$\\
        Fiducial frequency derivative* & $\dot{f}_0$ & $\mathrm{Hz}\,\mathrm{s}^{-1}$  & $-6.03824 \times 10^{-14}$ & $-8.847 \times 10^{-12}$\\
        Frequency deviation & $[f_-, f_+]$ & Hz & $[-3, 250] \times 10^{-7}$ & $[-3, 250] \times 10^{-7}$ \\
        Frequency derivative deviation & $[\dot{f}_-, \dot{f}_+]$ & $\mathrm{Hz}\,\mathrm{s}^{-1}$ & $[-6.04, 6.04] \times 10^{-15}$ & $[-4.9, 5.1] \times 10^{-14}$\\
        Frequency bin size & $\eta_f$ & Hz & $1.7 \times 10^{-8}$ & $1.7 \times 10^{-8}$ \\
        Frequency derivative bin size & $\eta_{\dot{f}}$ & $\mathrm{Hz}\,\mathrm{s}^{-1}$ & $1.21 \times 10^{-15}$ & $2 \times 10^{-15}$ \\
        Timing noise strength & $\sigma$ & $\mathrm{Hz}\,\mathrm{s}^{-3/2}$ & $2.56 \times 10^{-18}$ & $2.05 \times 10^{-18}$ \\
        Mean ToA uncertainty* & $\langle\sigma_\mathrm{ToA}\rangle$ & ms & 5.327 & $0.506$ \\
        Bayes factor threshold & $K_\mathrm{th}$ & None & $10^{1/2}$ & $10^{1/2}$\\
        \hline
    \end{tabular}
    \label{tab:hmm_params_J1452-6036}
\end{table*}

\bsp	
\label{lastpage}
\end{document}